\documentclass{sig-alternate}
\usepackage{graphicx}
\usepackage{multirow}
\usepackage{url}

\begin{document}
%
\conferenceinfo{WOODSTOCK}{'97 El Paso, Texas USA}

\title{Engineering Adaptive Digital Investigations\\ using Forensic Requirements\titlenote{
We acknowledge SFI for grant 10/CE/I1855 and ERC Adavanced Grant
(ASAP) no. 291652. We also thank Pavel Gladyshev and Mark McGloin for
useful discussions.}}

%
%
%
%
%

\author{
%
%
Liliana Pasquale$^{1}$, Yijun Yu$^{2}$, Luca Cavallaro$^{1}$\\
 Mazeiar Salehie$^{1}$, Thein Than Tun$^{2}$, Bashar Nuseibeh$^{1,2}$ \\
\affaddr{$^{1}$Lero - the Irish Software Engineering Research Centre,
  University of Limerick, Ireland}\\
\affaddr{$^{2}$Department of Computing and Communications, The Open University, UK}
}

\maketitle
\begin{abstract}
  A digital forensic investigation aims to collect and analyse the
  evidence necessary to demonstrate a potential hypothesis of a
  digital crime. Despite the availability of
  several digital forensics tools,
  investigators still approach each crime case from scratch,
  postulating potential hypotheses and analysing large volumes of
  data.  This paper proposes to explicitly model {\em forensic
    requirements} in order to engineer software systems that are
  forensic-ready and guide the activities of a digital investigation. Forensic
  requirements relate some speculative hypotheses of a crime to the
  evidence that should be collected and analysed in a crime scene. In
  contrast to existing approaches, we propose to perform  {\em proactive
    activities} to preserve
  important - potentially ephemeral - evidence, depending on the risk of a
  crime to take place. Once an investigation starts,
  the evidence collected proactively is analysed to assess if
  some of the speculative hypotheses of a crime hold and what further
  evidence is necessary to support them. For each hypothesis that is
  satisfied, a
  structured {\em argument} is generated to demonstrate how the evidence
  collected supports that hypothesis. 
Our evaluation results suggest that the approach provides 
correct investigative findings and reduces
  significantly the amount of
  evidence to be collected and the hypotheses to be analysed.
\end{abstract}




\section{Introduction}
\label{sec:intro}

A digital forensic investigation~\cite{Palmer.DFRWS.2001} aims to
collect and analyse the evidence necessary to demonstrate a potential
hypothesis of a digital crime, which explains how that crime was committed,
what harm was done, and who was responsible. A number of
processes~\cite{Carrier.DFRW.2004,Pollitt.DF.2007,Beebe.DI.2005} and
several digital forensics
tools~\cite{SpectorSoft,ArxSys,GuidanceSoftware,Sleuthkit,FTK,IEF,Helix}
can be used to conduct a digital investigation. However,
digital investigations remain highly human-intensive, and
investigators usually approach each crime case from scratch, by
postulating potential hypotheses and manually analysing large volumes
of -- often irrelevant -- data. Existing tools do not provide any
investigative direction to suggest the potential hypotheses, including
the evidence they require to be demonstrated, their likelihood of
being true, or the evidence necessary to demonstrate them. Some
evidence can also be ephemeral, as it can be concealed by an attacker
or it can come from volatile sources. Indeed this evidence 
might be lost if it is not preserved before an investigation starts.

A software engineering challenge in digital forensics is to build
systems that are forensic-ready~\cite{Rowlingson.IJDE.2004}, which
maximises the potential to use digital evidence whilst minimising the
costs of an investigation. Such systems not only should preserve all
necessary evidence, but they should also help investigators assess the
likelihood of potential hypotheses, and link the evidence collected to
the findings of an investigation.
 To address this challenge, this
paper proposes to explicitly model {\em forensic requirements} that
relate some speculative hypotheses of a crime to the evidence to be
collected in the crime scene. The hypotheses express how a crime can
be committed and are formalised as logic expressions.  The crime scene
represents a bounded environment where a crime can be perpetrated.
Forensic requirements also capture patterns of ``suspicious events''
that indicate that a crime may be taking place and for which all
necessary evidence should be collected for some time.

Forensic requirements are used to configure a
digital forensics process that includes proactive and reactive
activities.  {\em Proactive activities} -- evidence collection and analysis
-- are performed during the normal system operation. Proactive collection preserves
important -- potentially {\em ephemeral} -- evidence, which might otherwise
be lost, before an investigation starts. Proactive analysis detects ``suspicious events'' and therefore
enables the collection of additional evidence for some time. 
Once an investigation starts, some {\em reactive activities} are
performed. First, forensics requirements are used to identify all
potential 
hypotheses of a crime. 
Hypotheses are formalised in the Event
Calculus~\cite{Mueller.2006} and are examined  using {\em Decreasoner}~\cite{Decreasoner}, an Event
Calculus analyser. The
findings of this preliminary analysis are presented to the
investigator, suggesting which hypotheses should be investigated
(because they are likely) or ignored (because they have been
formally refuted). For each hypothesis that may hold, our process
suggests the remaining evidence that needs to be collected, and
reactively re-analyses the crime in light of the new evidence. When a
hypothesis is fully satisfied, our process automatically generates a
structured {\em argument}~\cite{Toulmin.2003} that demonstrates how the evidence
collected formally supports that hypothesis. 

This paper is a first step towards engineering forensic-ready
systems and it builds on the assumption
that all potential crimes are known and can be specified  in
advance. The paper provides three main contributions. First, it proposes a
novel adaptive digital forensics process - initially sketched
in~\cite{Pasquale.RE.2013} - to systematically perform the activities
to be conducted before and during a
digital investigation. The process
performs proactive activities to preserve important evidence and
suggests immediate investigative directions. Second, the paper
introduces the notion of forensic requirements to systematically
configure the activities of the proposed digital forensics process depending on a
specific crime scene and on the potential hypotheses of a crime.
Finally, the paper explains how structured arguments can be used to present
the findings of an investigation. We illustrate our approach on a criminal
case of confidentiality infringement and evaluated it on a
  realistic digital forensic scenario~\cite{Garfinkel.DFRWS.2009}. Our
  results suggest that the approach provides correct
  investigative findings and reduces significantly the amount of
  evidence to be collected and the number of hypotheses to be analysed compared to
  traditional digital investigations. 

  The rest of the paper is organised as follows. Section II presents
  the background and related research, while Section III illustrates a
  working example. Section IV provides an overview of our proposed
  digital forensics process. Sections V, VI and VII use our example to
  illustrate forensics requirements, and the proactive and reactive
  activities of our digital forensics process, respectively. Section
  VIII discusses our evaluation results, Section IX reviews related
  work, and Section X concludes.

\section{Preliminaries}
\label{sec:preliminaries}

This Section clarifies the differences of our approach with
intrusion detection systems (IDS), illustrates the use of arguments, and
introduces the Event Calculus.

\subsection{Intrusion Detection Systems}

Our approach augments traditional digital investigations with
pro\-active activities, which are similar to the operations of
IDS~\cite{axelsson2000intrusion}. These detect and diagnose different kinds of attacks,
such as intrusions, DDoS, and application-level attacks, by collecting
and analysing data (network traffic) in real time. IDS perform
proactive collection and analysis to recognize specific patterns in
data that may suggest that an attack is taking place, discover new
attacks, and re-configure firewall rules to counteract the
attacks. However, IDS are not primarily designed to gather
forensically meaningful evidence necessary to explain a crime. In
contrast, the objective of this work is to engineer systems that are
forensic-ready and are able to perform proactive evidence collection
and analysis respectively to identify suspicious events and preserve additional
evidence necessary to reconstruct a crime after it has been committed.

\subsection{Argumentation}

To explain the findings of an investigation we use a form of argument inspired by the work of Toulmin~\cite{Toulmin.2003}. Toulmin-style arguments capture  relationships between a {\em claim} and domain properties ({\em grounds} and {\em facts}), the assumptions that eventually support the grounds ({\em warrants}), and the reasons why the argument might not be valid ({\em rebuttals}). Arguments have been applied successfully to model and analyse privacy and security requirements. Haley et al.~\cite{Haley.TSE.2008}, have used Toulmin arguments to recursively represent the rebuttals and mitigations when reasoning about the satisfaction of security requirements. In their approach, security requirements are expressed as claims that are supported by grounds and warrants. Rebuttals show evidence that contradicts other arguments, whilst mitigations describe how rebuttals may be avoided or tolerated. Franqueira et al.~\cite{Franqueira.RE.2011} combine security arguments with risk assessment exploiting publicly available security catalogues. Privacy arguments~\cite{Tun.RE.2012} have been used to analyse selective disclosure requirements allowing reasoning about the satisfaction of personal non-disclosure requirements when context changes. 

Our work uses arguments to link the hypotheses of a crime (claims) to the evidence to be collected (facts), and the properties (grounds) and the assumptions (warrants) that hold in the crime scene. Since all these elements are represented explicitly in advance as forensic requirements, when a hypothesis is satisfied, it is possible to build an argument systematically to explain how a crime took place depending on the evidence collected. The derived argument structure can help provide robust evidence to support a hypothesis.

\subsection{Event Calculus}

To facilitate formal reasoning, the hypotheses of a crime are
expressed in the Event Calculus, a language based on first-order
predicate calculus. The Event Calculus is well suited to describing
and reasoning about event-based temporal systems~\cite{Giannakopoulou.2003}. The calculus relates
event sequences to {\em fluents} that denote the states of a
system. Fluents are initiated by an event and cease to hold when
terminated by another event. Event claulcus also includes non-temporal
predicates and functions that return a non-boolean result. Table~\ref{tab:I} gives the meanings of the elementary predicates and quantifiers of the subset of the calculus we use in this paper. The Event Calculus supports both {\em deductive} and {\em abductive} reasoning. Deduction uses the description of the system behaviour together with the history of events occurring in the system to derive the events/fluents that occur/hold at a particular point in time. Abduction determines the sequence of events that must have occurred to allow a set of events/fluents to hold/occur.

\begin{table}\centering
\caption{Event Calculus Predicates and Quantifiers}
\begin{small}
\begin{tabular}{|l|l|}\hline
\bf Predicate	& \bf Meaning \\\hline\hline
$\mbox{Happens}(a,t)$	& Action $a$ occurs at time $t$ \\\hline
$\mbox{Initiates}(a,f,t)$	& Fluent $f$ starts to hold after action $a$ at time $t$ \\\hline
$\mbox{Terminates}(a,f,t)$	& Fluent $f$ ceases to hold after action $a$ at time $t$ \\\hline
$\mbox{HoldsAt}(f,t)$	& Fluent $f$ holds at time $t$ \\\hline
$t_1 < t_2$	& Time point $t_1$ is before time point $t_2$ \\\hline\hline
\bf Quantifier	 & \bf Meaning \\\hline\hline
$\{\mbox{arg}_1,\ldots,\mbox{arg}_N\}$	& Exists: $\exists\ \mbox{arg}_1,\ldots, \mbox{arg}_N$ \\\hline
$[\mbox{arg}_1,\ldots,\mbox{arg}_N]$	& Forall: $\forall\ \mbox{arg}_1, \ldots, \mbox{arg}_N$ \\\hline
\end{tabular}
\end{small}
\label{tab:I}
\end{table}

The Event Calculus is a suitable formalism for digital investigations~\cite{Willassen.SAC.2008}. Events can represent the actions that occur in a crime scene, which can also be perpetrated by an offender. Fluents can represent the state of the elements modelled in the crime scene. Deductive reasoning can be used for proactive analysis to determine if a suspicious event happened. Abductive reasoning can be used for reactive analysis to speculate on what sequence of events might have occurred, if any, to allow the system to satisfy the condition claimed in the hypothesis. 

\def\Alice{{\tt Alice}}
\def\Bob{{\tt Bob}}
\def\T225{{\tt T225}}
\def\Doc{{\tt Doc}}
\def\Mone{{\tt M1}}
\def\Mtwo{{\tt M2}}
\def\Mthree{{\tt M3}}
\def\CCTV{{\tt CCTV camera}}
\def\NFCreader{{\tt NFC reader}}
\def\USBpen{{\tt USB pen}}
\def\Enter{{\tt Enter}}

\section{Working Example}
\label{sec:working-example}

Although enterprises often invest significant resources to develop incident response plans, very little effort is devoted to the identification and preservation of digital evidence and the structuring of processes for possible prosecution. The need for enterprises to prepare themselves for a crime investigation has also been highlighted in ISO27001/2~\cite{ISO}. This can be particularly useful for those crimes following a specific pattern and whose hypotheses can be modelled with little effort. For example, the FBI Financial Crimes Report of 2010--2011~\cite{FBICrimeReport} identifies the crime patterns in corporate and health care cases. Corporate fraud crimes often involve insider trading using confidential information. For example, in the Galleon Group case~\cite{NewYorkTimes}, insiders were charged for the unauthorized release of proprietary corporate information. While for the WikiLeaks case, an insider copied corporate sensitive information onto a CD. 

We assume that the crime scene is within a bounded environment, where crimes follow a pre-determined pattern. Inspired by the cases above, our working example is set in an enterprise building, where two employees (\Alice\ and \Bob) work. A sensitive document (\Doc) is stored on a server machine located in the office \texttt{T225}. Both \Alice\ and \Bob\ are authorized to access \texttt{T225} and to log on the machine \Mone.  Access to \texttt{T225} 
is controlled by a \NFCreader\ and is monitored by a \CCTV. \Bob\ and \Alice\ work on their personal desktops (\Mtwo\ and \Mthree, respectively), provided by the company. We also assume that potential offenders are the enterprise employees who can leak the sensitive document.

\section{Digital Forensic Process}
\label{sec:process}

\begin{figure}[htpb]
  \centering
  \includegraphics[width=.95\columnwidth]{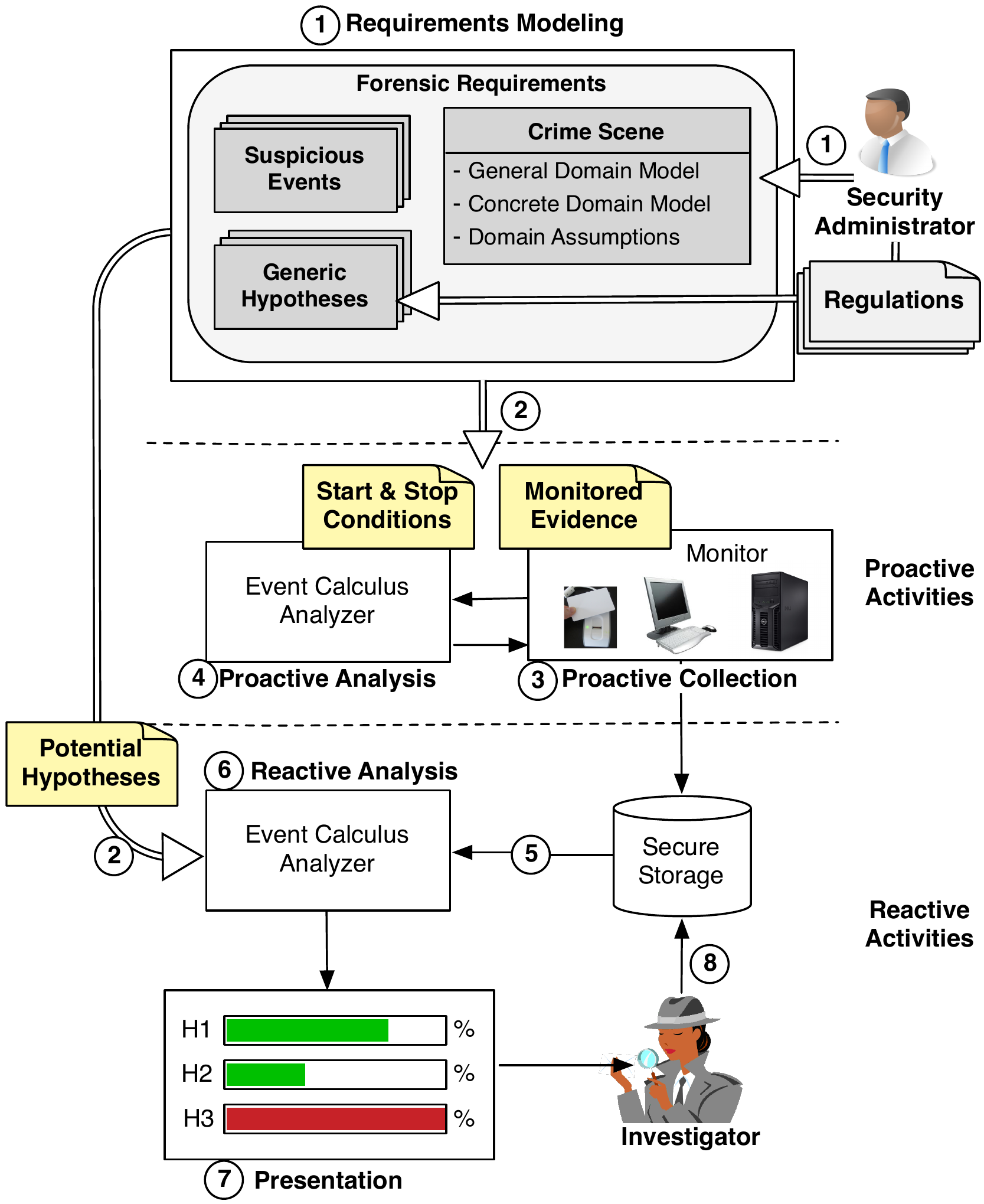}
  \caption{Adaptive Digital Forensics Process.}
  \label{fig:overall}
\end{figure}

Figure~\ref{fig:overall} provides a schematic overview of our digital
forensics process, comprising eight steps that we now describe.

\textbf{1) Requirements Modelling}. Forensic requirements represent
the crime scene and the speculative hypotheses of a crime. A security
administrator, who has relevant domain expertise, model the crime
scene, which describes a bounded environment where a crime can be
perpetrated. It includes a general and a concrete domain model. The
general domain model represents the generic entities (e.g., employees,
files, locations) and evidence sources (e.g., computers, cameras),
including their possible states. Evidence sources are those elements
from which evidence can be collected. The concrete domain model
instantiates the general model on the concrete elements available in a
crime scene, and identifies their initial state and the events to be
monitored. A concrete domain model for our example can include
company's employees (\Alice\ and \Bob), a confidential document
(\Doc), computers (\Mone, \Mtwo, and \Mthree) and an office (\T225).
As an initial state, one can assume that \Doc\ is stored in \Mone,
which in turn is located in \T225. The crime scene also includes a set
of assumptions about the domain.

In this work a crime is conceived as non-compliance to a governmental or
corporate regulation~\cite{Ingolfo.REFSQ.2013} and the hypotheses
of a crime represent how  a violation of these regulations can happen. For
example, if a corporate regulation states that a confidential
document should not exfiltrate from the PC where it is stored,
a generic hypothesis can express that an employee is
logged onto the PC where the document is stored, s/he is in the
room where the PC is located, and copies the document onto a USB storage.  

For each generic hypothesis of a crime, a suspicious event is also
represented. This expresses a condition in the crime scene that might
indicate that the corresponding hypothesis of a crime is likely to be
satisfied, and all the necessary evidence to prove it must be
collected in advance until the suspicious event condition no longer holds. An example of
suspicious event may indicate that a user is logged onto the machine
in which the document is stored and mounts a USB storage.

\textbf{2) Configuration}. Forensic
requirements are used to configure the proactive and reactive activities of a
digital forensic process. Suspicious events are used to configure the
conditions that will be checked by the Proactive Analysis to start/stop the
full evidence collection (collection of all possible and necessary
evidence to satisfy a hypothesis). Forensic requirements are also used to generate the potential
hypotheses given as input to the Reactive
Analysis. Potential hypotheses are defined by instantiating the
generic hypotheses on the concrete elements of the crime
scene. For our example, a number of hypotheses ($3^3=27$) will be
generated depending on who can be in \T225, who can log on \Mone, and who
can own the storage device which should be mounted on \Mone\ (\Alice\ only, \Bob\ only, or both \Alice\ and \Bob).

\textbf{3) Proactive Collection.} If no start condition is satisfied,
the Proactive Collection only gathers all the events necessary to
verify whether any of the start
conditions holds. Otherwise, it will
additionally gather all possible and necessary events to demonstrate
the hypotheses associated with the start conditions that
hold at that time. The Proactive Collection stores all monitored events
securely and sends to the  Proactive Analysis only those events
 necessary to verify the start and stop conditions.

\textbf{4) Proactive Analysis.} When new evidence is available, the
Event Calculus Analyzer (Analyzer) checks if the conditions to start/stop the full evidence
collection are satisfied. In case a start condition is satisfied, the
Proactive Analysis signals the Proactive Collection to enable the full
evidence collection for the hypothesis associated
with the satisfied start condition. When a stop condition is satisfied, the
Proactive Analysis signals the Proactive Collection to gather only the
evidence necessary to verify the corresponding start condition.

\textbf{5) Investigation Set-up.} Once an investigation has started,
the Reactive Analysis retrieves the data collected proactively from the Secure Storage.

\textbf{6) Reactive Analysis.} The Event Caluclus Analyzer (Analyzer) evaluates the satisfaction
of each potential hypothesis and sends the results to the Presentation
activity. In particular, it identifies the hypothesis that can still
hold and cannot be refuted on the base of the data collected.
For example,  we can assume that the events retrieved from the Secure Storage make
 the Reactive Analysis conclude that \Bob\ was the only user logged on
 \Mone\ who copied the \Doc\  on a \USBpen . In this case, only those
 hypotheses that claim that \Bob\ was the only employee logged on
 \Mone\ who copied the \Doc\ on a \USBpen\ can still hold.

\textbf{7) Presentation.} This activity shows the satisfaction of each potential hypothesis. The investigator selects the hypotheses s/he wants to focus on and receives indications regarding the remaining evidence to be collected. For example, the investigator is suggested to collect additional evidence from a camera to confirm that \Bob\ was in \T225\ when he logged on \Mone, and from \Bob's computer to verify whether \Bob\ owns the \USBpen. 

\textbf{8) Reactive Collection.} The investigator retrieves additional
evidence, by using existing commercial tools (e.g.,
Sleuthkit~\cite{Sleuthkit}) and stores it securely. The
cycle (activities 5-8) continues until the investigator identifies
a hypothesis that is fully satisfied. In this case, the Presentation automatically
generates an argument that demonstrates how the evidence collected
formally supports it.

\section{Forensic Requirements}
\label{sec:forensicRequirements}

\def\In{{\tt In}}
\def\Possess{{\tt Possess}}
\def\Location{{\tt Location}}
\def\Logged{{\tt Logged}}
\def\Login{{\tt Login}}
\def\Logout{{\tt Logout}}
\def\Stored{{\tt Stored}}
\def\Mounted{{\tt Mounted}}
\def\PlacedIn{{\tt PlacedIn}}
\def\AccessToFile{{\tt AccessToFile}}

Forensics requirements represent the crime scene, the hypotheses of a crime and suspicious events.

\subsection{Crime Scene}
As described previously, a crime scene consists of a general and a concrete domain model and a set of domain assumptions.

\begin{figure}[htpb]
\centering
\includegraphics[width=0.9\columnwidth]{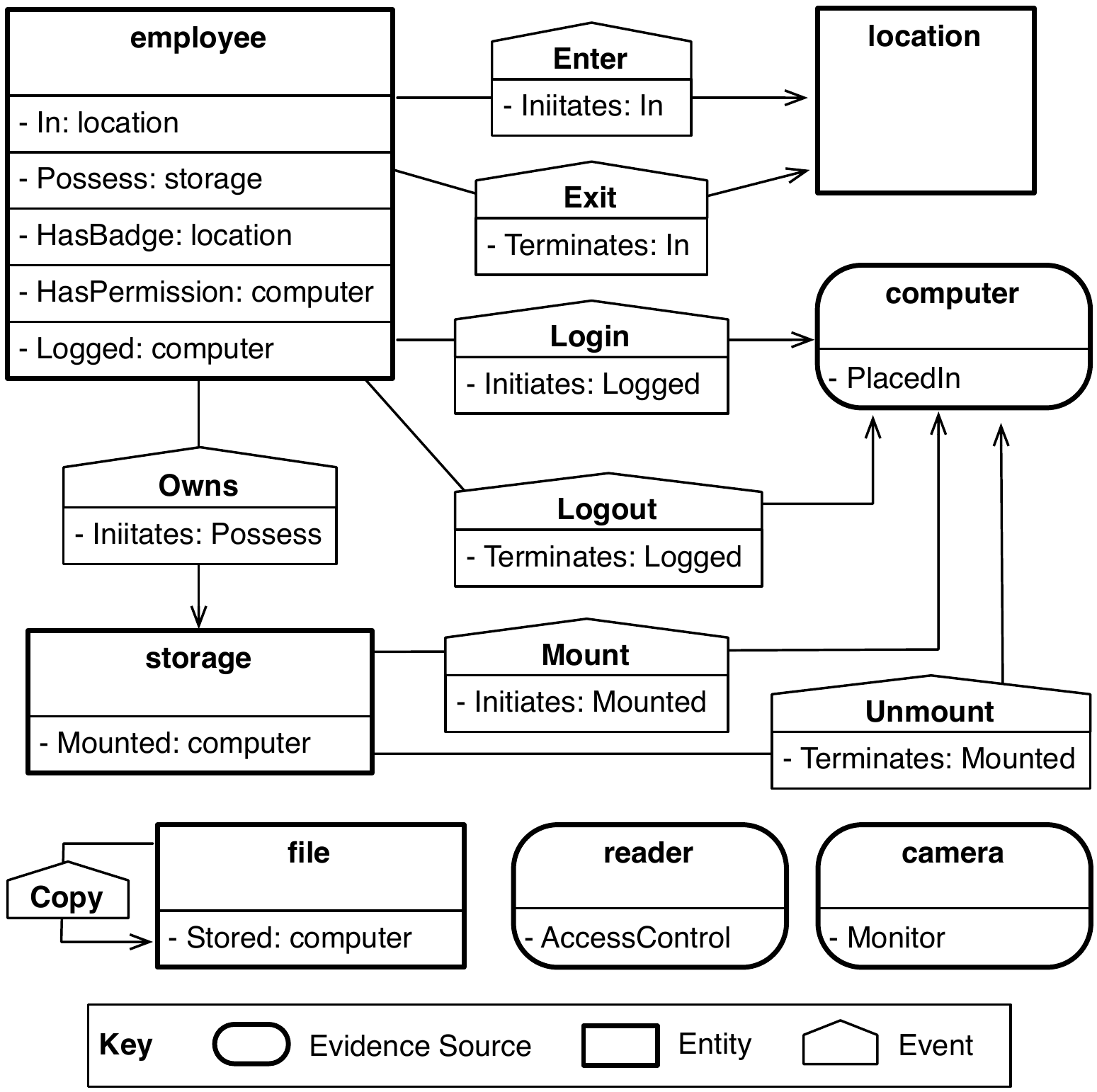}
\caption{The General Domain Model of our example.}\label{fig:2}
\end{figure}

{\em 1) A General Domain Model} represents the ``type'' of the entities and evidence sources in the crime scene. The general domain model of our example is shown in Figure~\ref{fig:2}. Entities can be {\em employees}, {\em files}, {\em storage devices}, and {\em locations}, while evidence sources can be {\em computers}, {\em cameras}, and {\em card readers}. Entities and evidence sources can be associated with some states, represented as attributes in Figure~\ref{fig:2}. For example, an employee can be \In\ a location and \Logged\ on a computer. S/he can also {\tt Possess} a storage device, and be authorized to log on a computer ({\tt HasPermission}) or to access a location ({\tt HasBadge}). A file and a storage device can be respectively \Stored\ and \Mounted\ on a computer. A reader and a camera can monitor the access to a location (states {\tt AccessControl} and {\tt Monitor}, respectively), and a computer can be in a location (\PlacedIn). The general domain model also includes general events. An event must have a source (arrow's tail) and an observer (arrow's head). Events can initiate or terminate one of the states of their source, and also have additional parameters. For example, \Login\ and \Logout\ respectively initiate and terminate state \Logged\ of an employee for the computer on which the login/logout operation was performed. 

\begin{figure}[htpb]
\centering
\includegraphics[width=0.9\columnwidth]{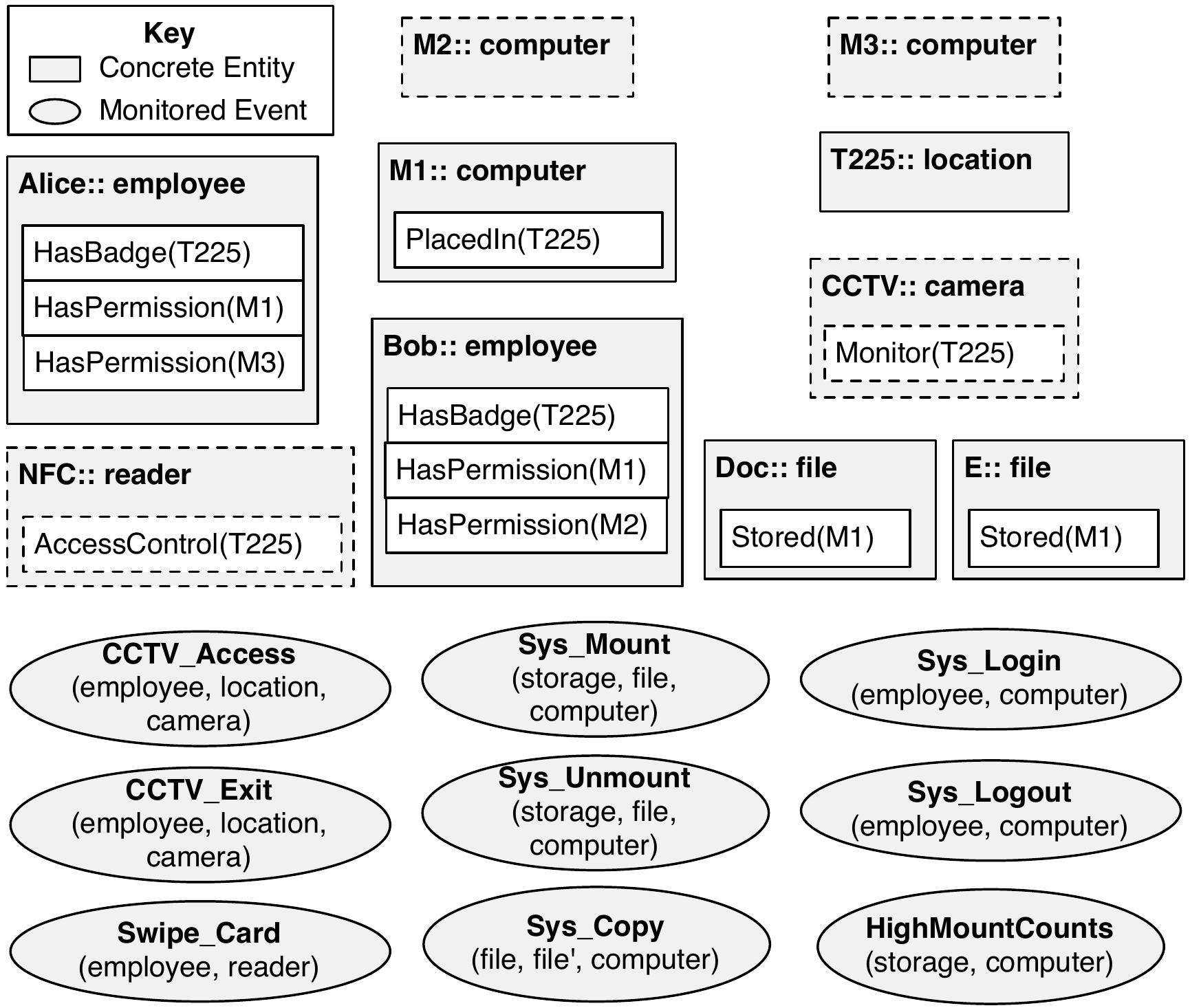}
\caption{The Concrete Domain Model of our example.}\label{fig:3}
\end{figure}

{\em 2) A Concrete Domain Model} instantiates the entities and the evidence sources specified in the general model depending on the concrete elements that are present in the crime scene. As shown in Figure~\ref{fig:3}, the concrete domain model of our example includes a location (room \T225), three computers (\Mone, \Mtwo, \Mthree), two employees (\Alice\ and \Bob), a camera ({\tt CCTV}), a NFC reader (\NFCreader), and a confidential file (\Doc). The concrete domain model also devises the initial states for some elements of the crime scene. For example, \Doc\ is stored on \Mone, which in turn is located in \T225. {\tt CCTV} and {\tt NFC} control accesses to \T225. \Alice\ and \Bob\ are authorized to access \T225, and have permission to log on \Mone\ and their own corporate computers (\Mthree\ and \Mtwo, respectively). 

The concrete domain model also represents monitored events identifying the concrete evidence that can be collected from a crime scene. The first and the last parameter of a monitored event identify respectively its source and observer -- the evidence source from which an event must be collected. For example, {\tt Sys\_Login} and {\tt Sys\_Logout} respectively identify login and logout operations performed by an employee (source) on a computer (observer).\\ {\tt Swipe\_Card} and {\tt HighMountConts} signal respectively that an employee swiped his/her card on a reader and a storage device was mounted several times (e.g., more than 3 times) on a computer. Any other middle parameter of an event -- if present -- identifies additional information. For example, {\tt Sys\_Copy} signals that a file was copied onto a directory ({\tt file'}). The data coming from the sources of evidence represented with dashed lines cannot be collected proactively, because they cannot be gathered automatically, such as the evidence from a {\tt CCTV} ({\tt CCTV\_Access} and {\tt CCTV\_Exit}), or because it is illegal to do so without a search warrant, such as evidence from an employee's corporate computer (e.g., {\tt HighMountCounts} from \texttt{M2} or \texttt{M3}).

\begin{figure*}[htpb]
\centering
\includegraphics[width=1.9\columnwidth]{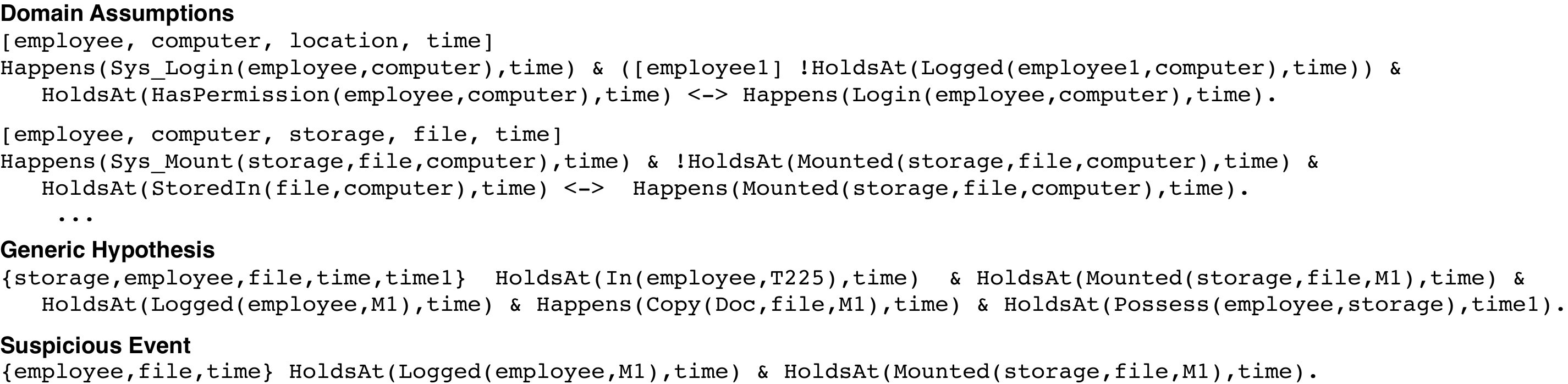}
\caption{Domain assumptions, generic hypothesis, and suspicious event of our example.}\label{fig:4}
\end{figure*}

{\em 3) Domain Assumptions} are inference rules that link monitored events to the events and the states represented in the general domain model. Figure~\ref{fig:4} shows some domain assumptions of our example. They are represented as implications or equivalences among Event Calculus predicates expressions. The first domain assumption states that an employee logins on a computer (\Login) if and only if s/he has the necessary permission, a monitored event {\tt Sys\_Login} signals that s/he performed the login, and no user is already logged on that computer. Note that the \Login\ event initiates state \Logged\ of the aforementioned employee. The second domain assumption states that a storage device is mounted at a specific mount point ({\tt file}) in a \texttt{computer}, if and only if the storage is not already mounted, the mount point exists on that computer ({\tt StoredIn}), and  an attempt to mount a  storage device was performed ({\tt Sys\_Mount}).

\subsection{Hypotheses and Suspicious Events}

The generic hypothesis of a crime and the suspicious event of our
example are represented in Figure~\ref{fig:4}. They are formalized as
an Event Calculus expression on the events and the states of the
elements in the crime scene. Our hypothesis states that at least an
employee is in \T225; one of the employees is logged on \Mone\ and
copies the \Doc\ on his/her \USBpen\, which is mounted on \Mone. A
suspicious event indicates that an employee is logged on \Mone\ and
mounts a \USBpen. Note also that suspicious event conditions can only
be expressed on those events that can be collected automatically
(i.e., proactively). A hypothesis of a crime (H) must always imply its
corresponding suspicious event condition (SE), but not vice-versa:
\begin{small}
  $H \rightarrow SE~\land \lnot(SE \rightarrow H)$.
\end{small}

Since proactive analysis is performed during the normal system
operations, we aim to reduce its complexity by keeping a suspicious
event condition simple (i.e., with a smaller number of predicates
compared to those included in its corresponding hypothesis). However,
on the other hand, the choice of a suspicious event condition should
still allow reducing the amount of evidence collected proactively, and
indeed it cannot be too simple and must still indicate if there is the
risk of a crime to happen. Note that a suspicious event condition can
only be specified if a hypothesis of a crime is composed of at least
two predicates.


\section{Proactive Activities}
\label{sec:proactive}

Proactive activities aim to collect evidence necessary to explain a
crime. Full evidence collection is performed only when a suspicious
event condition suggests that a crime is taking place, and is
terminated after such condition no longer holds.

\begin{figure}[htpb]
\centering
\includegraphics[width=0.95\columnwidth]{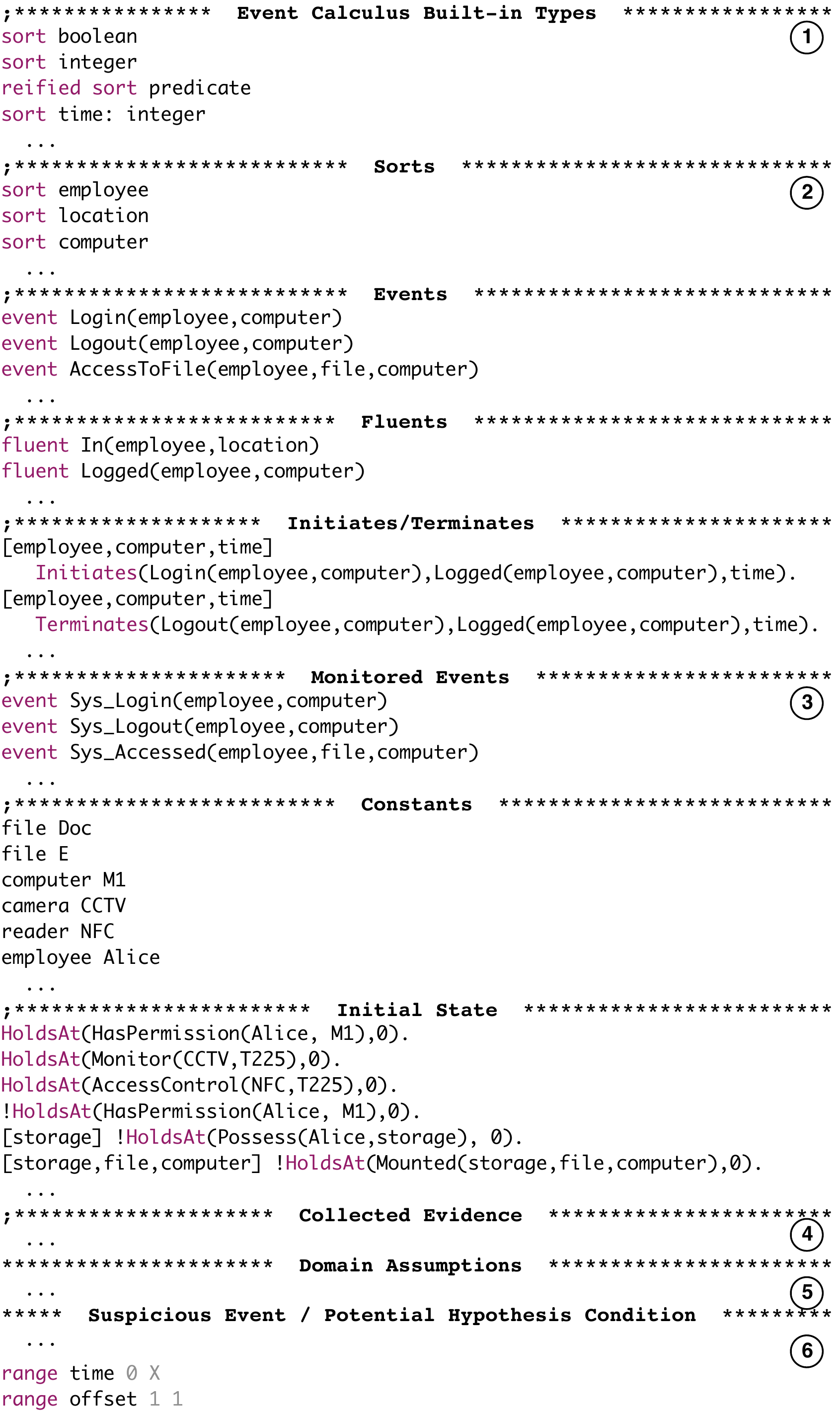}
\caption{A partial example of the Event Calculus specification.}\label{fig:5}
\end{figure}

The Proactive Analysis is initially configured to check the start conditions necessary to
activate the full evidence collection. In particular, the
Event Calculus Analyzer (Analyzer) receives as input an Event Calculus specification, shown in Figure 5,
which is obtained from the forensic requirements. The first part of
the specification is static and is necessary to define a set of
built-in types ({\tt sort}), such as {\tt boolean, integer, time,
  predicate}, {\tt event}, and {\tt fluent}. The general domain model
is used to derive the second part of the specification. In particular,
entities and evidence sources are translated into a set of Event
Calculus types ({\tt sorts}), while events are translated into Event
Calculus {\tt events}. Source, additional parameters -- if present --
and observer of each event are also encoded as event parameters. For
each state an entity/evidence source can assume, a corresponding
fluent is created, having the state's subject and object as
parameters. An {\tt Initiates/Termi\-nates} predicate is generated for
each state initiated/terminated by an event. For example,
Figure~\ref{fig:5} shows how events \Login\ and \Logout\ respectively
initiate and terminate state \Logged.

The concrete domain model is used to generate the third part of the
Event Calculus specification. The initial states are used to identify the initial predicates, which hold at time 0 for some of the Event Calculus constants. For all the other possible states that do not hold at time 0, a negated predicate is created. Monitored events are translated into Event Calculus events. Concrete entities and evidence sources are translated into Event Calculus constants. 
The fourth part of the Event Calculus specification is generated from
the evidence gathered during the Proactive Collection. The fifth and
the sixth parts include respectively the domain assumptions and the
condition characterising a suspicious event. The last part also
indicates the time interval that should be considered during the
analysis (range), which in turn depends on the number of time instants
in which the Proactive Collection gathers some new event.

The Proactive Collection is initially configured to collect the
monitored events necessary to check the suspicious event
conditions. To this extent, it is necessary to identify the events and the entities' states a
suspicious event condition predicates on. The
suspicious event condition shown in Figure~\ref{fig:5} is specified
over state \Logged\ and event {\tt Mount}. Then, for each state, the
initiating and terminating events are identified, from the general
domain model. For example, events \Login\ and \Logout\ are identified
from state \Logged. Finally the monitored events to be collected to
detect these complex events are identified from the domain
assumptions. For our example, the Proactive Collection is configured
to collect events {\tt Sys\_Login}, {\tt Sys\_Logout}, {\tt
  Sys\_Mount}, and {\tt Sys\_Un\-mount} from \Mone\ to detect respectively complex
events \Login, \Logout, {\tt Mount}, and {\tt Unmount}.

When the Proactive Collection gathers some new events, these are sent
to the Analyzer that updates the Event Calculus specification to be
evaluated. In particular, it increments the time range and  adds to
the Collected Evidence (part 4 in Figure~\ref{fig:5}) a set of
predicates representing the events that took place in the last time
instant. For example, if event {\tt Sys\_Login(Bob,M1)} happened at
instant 5, the Collected Evidence part is updated with predicate {\tt
  Happens(Sys\_Login(Bob,M1),5)}. Furthermore, all other
possible ways in which event {\tt Sys\_Login} can take place are
negated (e.g., {\tt
  !Happens\-(Sys\_Login(Alice, M1),5)}).

When the start condition is satisfied, the Analyzer signals the
Proactive Collection to enable the full evidence collection. In
our example, the Proactive Collection will also gather all {\tt Sys\_Copy}
from \Mone. In the meanwhile the Analyzer controls if the  stop condition is
satisfied, and, in that case, it signals the Proactive Collection to
terminate the full evidence collection. The sequence of events - including their timestamps - collected
during two consecutive times in which a stop condition holds will be grouped into a
snapshot that is stored securely. Note that we assume that the stop
condition initially holds.

\section{Reactive Activities}
\label{sec:reactive}

\begin{figure}[htpb]
\centering
\includegraphics[width=0.9\columnwidth]{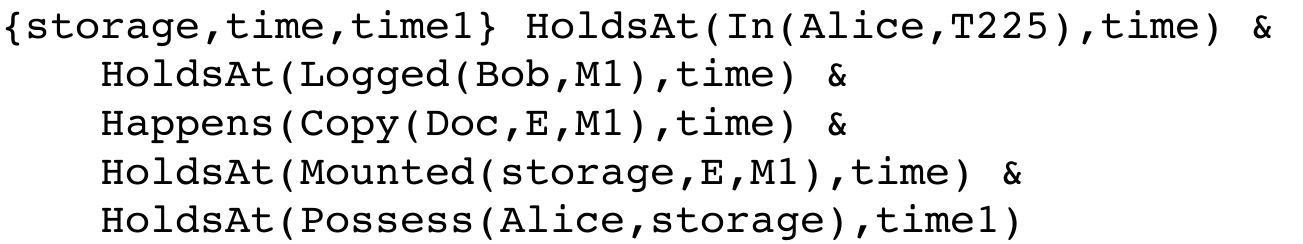}
\caption{Formalisation of Hypothesis H4.}\label{fig:6}
\end{figure}

Reactive Analysis is configured with the potential hypotheses of a crime which will be evaluated for all the events snapshots preserved proactively. These are obtained by instantiating the generic hypotheses depending on the concrete elements present in a crime scene. As shown in Table~\ref{tab:II}, for our example, 27 hypotheses will be generated depending on who can be in \T225, who can log on \Mone,  and who can own the storage device that is mounted on \Mone . The potential hypotheses are formalised in the Event Calculus, as shown in Figure~\ref{fig:5}. In this case, the Monitored Events (part 4) will include one of the events snapshots collected proactively, while the Conditions (part 6) will represent the specific condition associated with a potential hypothesis of a crime. For example, the condition associated with H4 is shown in Figure~\ref{fig:6} and expresses that Alice is in room \T225, \Bob\ is logged on \Mone\ and copies the \Doc\ on \Alice's storage device, which is mounted on \Mone.

  \begin{table}
   \caption{Potential Hypotheses}
\centering
\begin{footnotesize}
  \begin{tabular}{|c|l|l|l|}
    \hline
    \textbf{Name}	&\textbf{In T225}	& \textbf{Logged}		& \textbf{USB Owner} \\\hline
    \textbf{ H1}	&Alice	&Alice		&Alice \\\hline
    \textbf{H2}	&Alice 	&Alice		&Bob \\\hline
    \textbf{H3}	&Alice	&Alice		&Alice, Bob \\\hline
    \textbf{H4}	&Alice	&Bob		&Alice \\\hline
    \textbf{H5}	&Alice	&Bob		&Bob \\\hline
    \textbf{H6}	&Alice	&Bob		&Alice, Bob \\\hline
    \ldots & \ldots & \ldots & \ldots \\\hline
    \textbf{H13}	&Bob	&Bob		&Alice \\\hline
    \textbf{H14}	&Bob	&Bob		&Bob \\\hline
    \textbf{H15}	&Bob	&Bob		&Alice, Bob \\\hline
    \ldots & \ldots & \ldots & \ldots \\\hline
    \textbf{H22}	&Alice, Bob	&Bob		&Alice \\\hline
    \textbf{H23}	&Alice, Bob	&Bob		&Bob \\\hline
    \textbf{H24}	&Alice, Bob	&Bob		&Alice, Bob \\\hline
    \ldots & \ldots & \ldots & \ldots \\\hline
    \textbf{H27}	& Alice, Bob	& Alice, Bob  & Alice, Bob\\\hline
  \end{tabular}
\end{footnotesize}

 \label{tab:II}
  \end{table}

\begin{figure}[htpb]
\centering
\includegraphics[width=0.9\columnwidth]{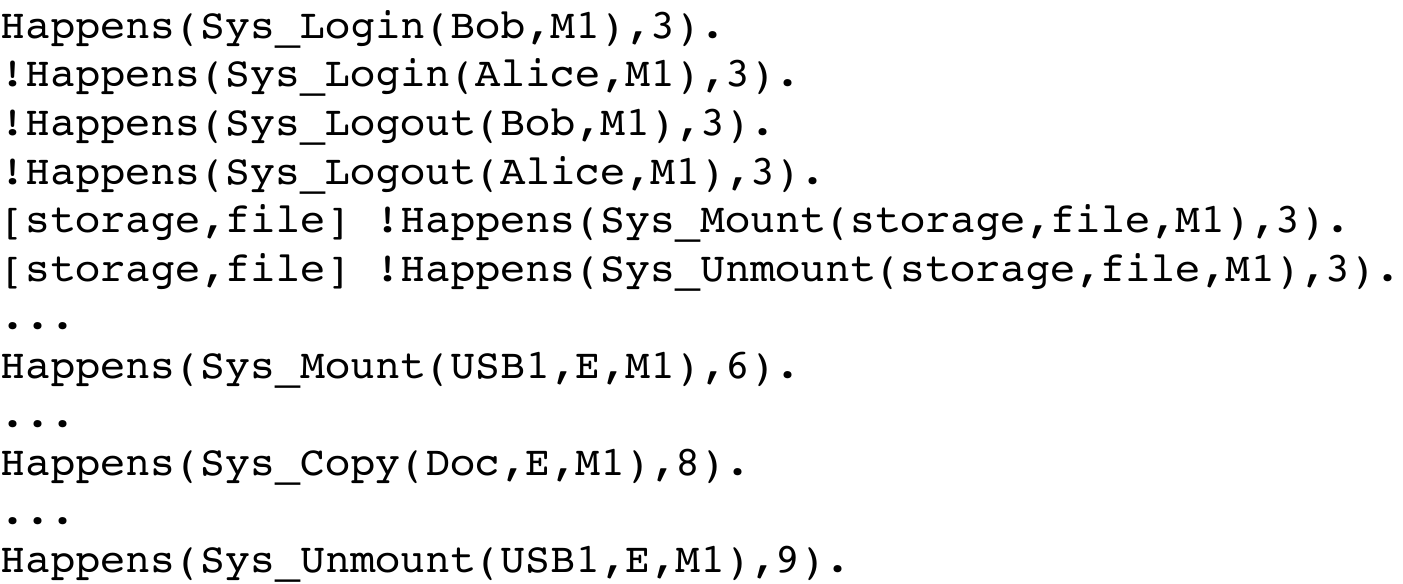}
\caption{Events Snapshot Example.}\label{fig:7}
\end{figure}

Once an investigation starts, the Analyzer  retrieves all the events snapshots that have been collected proactively and, for each of them, it checks if the potential hypotheses are  satisfied. Note that some of the events could not be monitored proactively, such as those that cannot be collected automatically (e.g. {\tt CCTV\_Access} from {\tt CCTV}). The Analyzer indeed performs abductive reasoning to verify if potential hypotheses can hold by speculating on the truth of these events. In other words, these events represent the missing evidence that must be collected to prove/refute the potential hypothesis.  For our example, we assume that the events snapshot on which the analysis is performed is the one represented in Figure~\ref{fig:7}. This indicates that \Bob\ was logged on \Mone\ and copied the \Doc\ on a \USBpen\ that was mounted on \Mone. For this events snapshot, the Analyzer discovers that only some of the potential hypotheses can still hold (H4-H6, H13-H15, and H22-H24).

\begin{figure}[htpb]
\centering
\includegraphics[width=0.7\columnwidth]{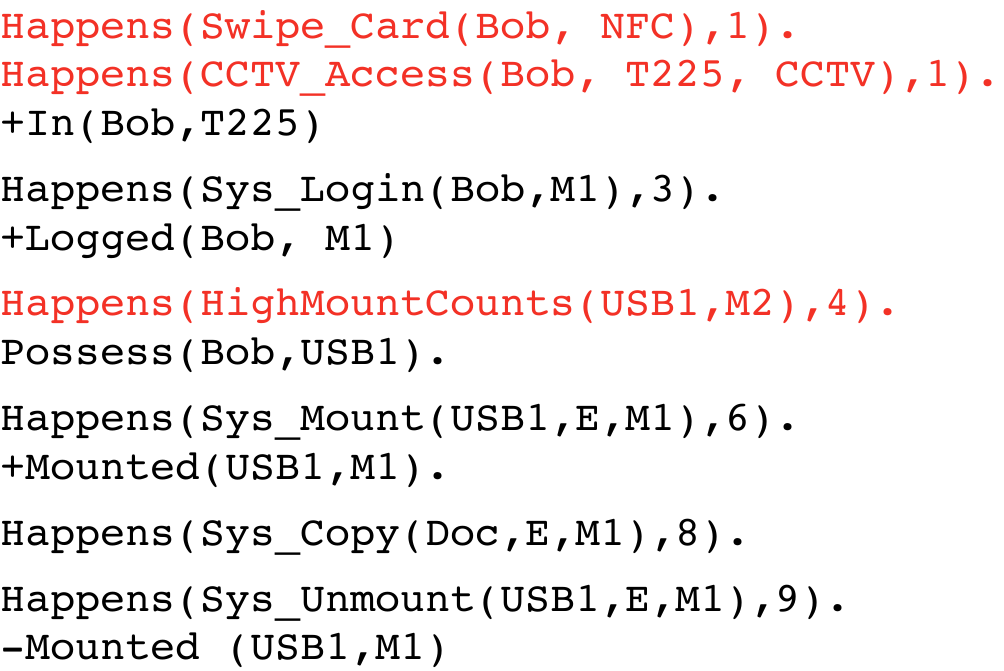}
\caption{Example of Additional Evidence}\label{fig:8}
\end{figure}

Analysis results are sent to the Presentation activity, which indicates the likelihood of each hypothesis depending on the proportion of predicates that have been demonstrated through the evidence collected (i.e., the predicates that have not been abducted~\cite{Freeman.IL.2008} and indeed are known).  For our example, the likelihood of hypotheses H4, H5, H13, and H14 is 60\%, since the predicates specified on fluents {\tt Logged}, {\tt Copy}, and {\tt Mounted} are satisfied by using monitored events, while the remaining predicates specified on fluents \In\ and \Possess\ leverage unkown events.  The likelihood of H6, H15, H22, and H23 is 50\%, as they include an additional predicate that verifies if \Alice\ and \Bob\ own the \USBpen\ that was mounted on \Mone\ (for H6 and H15) or are in room \T225\ (for H22 and H23). Finally the likelihood of H24 is about 42.86\%, since it includes two additional predicates that state that both \Bob\ and \Alice\ were in \T225\ and own the \USBpen\ that was mounted on \Mone. When an investigator selects a hypothesis s/he is willing to explore, the Presentation identifies additional evidence to be collected from the output of the Event Calculus analysis, by selecting those events that cannot be monitored automatically, as shown in Figure~\ref{fig:8}.  In case an investigator selects one of the most likely hypotheses, such as H5, the Presentation suggests collecting additional evidence (events {\tt Swipe\_Card} and {\tt CCTV\_Access}) from the {\tt NFC} and the {\tt CCTV} that control room \T225. \Bob's corporate computer (\Mtwo) should also be inspected to verify whether he mounted the \USBpen\ several times (event {\tt HighMountCounts}). 

\begin{figure*}[htpb]
\centering
\includegraphics[width=0.9\textwidth]{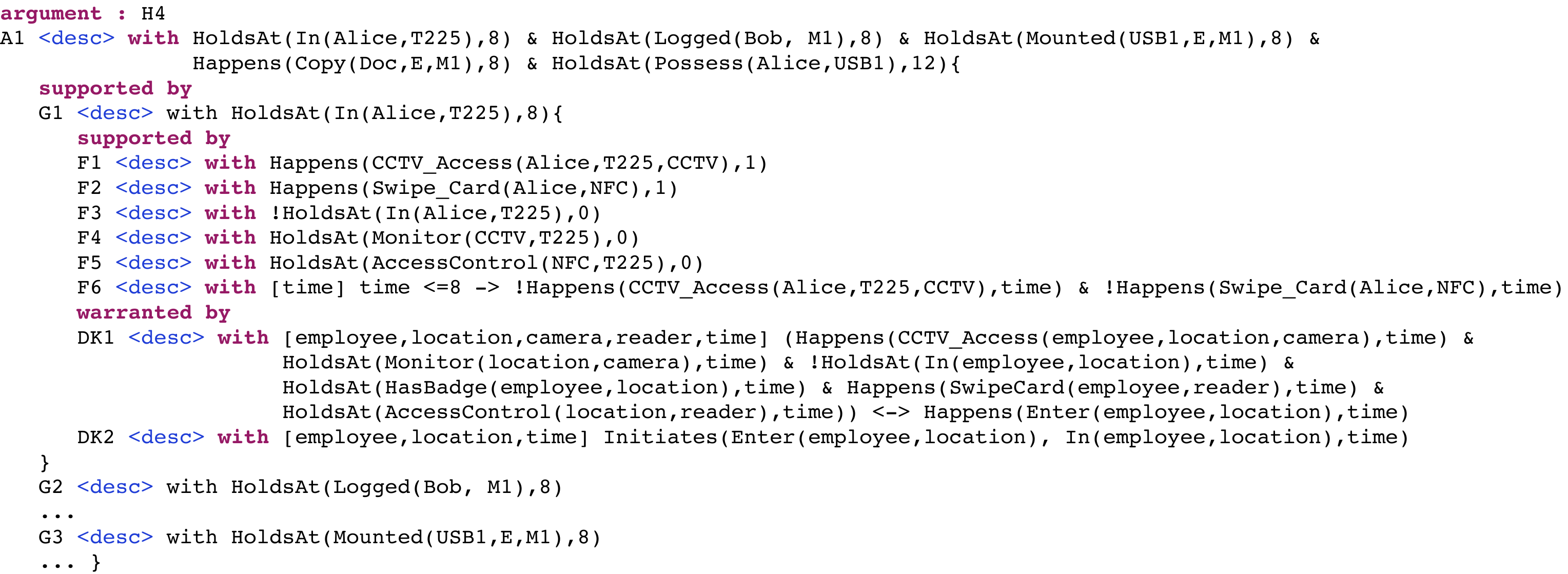}
\caption{Structured Argument that Explains Hypothesis H4.}\label{fig:9}
\end{figure*}

If an investigator collects events {\tt Happens(CCTV\_\-Access(\-Alice,T225),1)} and {\tt Happens(Swipe\_Card(Alice,NF\-C),1)}, it means that only \Alice\ was in \T225\ when \Bob\ logged in.  After the Reactive Analysis is re-performed, only 3 of the remaining hypotheses can still hold (H4-H6) with 80\% of likelihood for H4 and H5 and about 67\% of likelihood for H6. In case an investigator decides to explore H4, the Presentation suggests him/her to collect event {\tt  HighMountCounts(USB1,M3)} to justify that \Alice\ owns the USB storage. In case the investigator collects this evidence, only H4, and H6 can still hold as they will be respectively 100\%, and 83\% likely. At this stage, the investigator can continue to explore H6 that is still not completely satisfied.

Once a hypothesis is fully satisfied, the Presentation uses the analysis results and the representation of the forensic requirements to build a structured argument that explains how collected evidence demonstrates the claim made by that hypothesis. An argument is decomposed into grounds (sub-arguments)  associated with a predicate defined on an event or a fluent in the claim (hypothesis definition). A sketch of the argument generated for H4 is shown in Figure~\ref{fig:9}. This is decomposed into 5 grounds associated with each predicate of the claim. The predicate included in the grounds is further decomposed into lower-level grounds until only monitored events are identified (facts representing collected evidence). Facts and grounds in a sub-argument are related to their parent argument through domain assumptions (warrants). These explain how facts and grounds imply their parent argument. In Figure~\ref{fig:9}, ground {\tt G1} is associated with the predicate defined on the fluent \In. This ground is decomposed into facts that represent the monitored events that demonstrate that \Alice\ entered in \T225\ at instant 1 ({\tt F1}-{\tt F7}) and did not exit from the room till instant 8 ({\tt F8}). The domain assumptions that relate monitored events to state \In\ are {\tt DK1} and {\tt DK2}. {\tt DK1} relates state \In\ to the event that initiates it (\Enter), while {\tt DK2} relates event \Enter\ to facts {\tt F1}-{\tt F8} that trigger it.

\section{Evaluation}
\label{sec:evaluation}


Our evaluation was conducted on a public digital forensic
case~\cite{DigitalCorpora} where a confidential document of a startup
company  was
posted in the technical support forum of a competitor's website. The
document came from the computer of the CFO (Jean). The data-set
provides a copy of Jean's computer hard drive and a copy the
document.  The investigation aims to understand how did
the document get from Jean's computer to the competitor website. The
data-set is big enough to be realistic ($\sim$1.46GB), but it is small
enough to allow performing an investigation on a desktop computer.

Our evaluation assessed the correctness and effectiveness
of our approach compared to traditional digital investigations aided
by digital forensics tools. In particular, we verify if, under the assumptions that modeled hypotheses of a crime are
correct, our approach leads to the same - or even more accurate -
conclusions.
Effectiveness is estimated in terms of the amount of
evidence an investigator has to analyse that also affects the number
of hypotheses to be evaluated.
We conducted the same investigation by using available digital
forensics tools (Section 8.1), our approach
without proactive activities (Section~\ref{sec:react-invest}), and our
approach with proactive activities (Section~\ref{sec:case-3}). The
interested reader can find a detailed description of
the evaluation results in~\cite{Evaluation}.

\subsection{Traditional Digital Investigation}
\label{sec:8.2}

\begin{figure}[htpb]
\centering
\includegraphics[width=0.9\columnwidth]{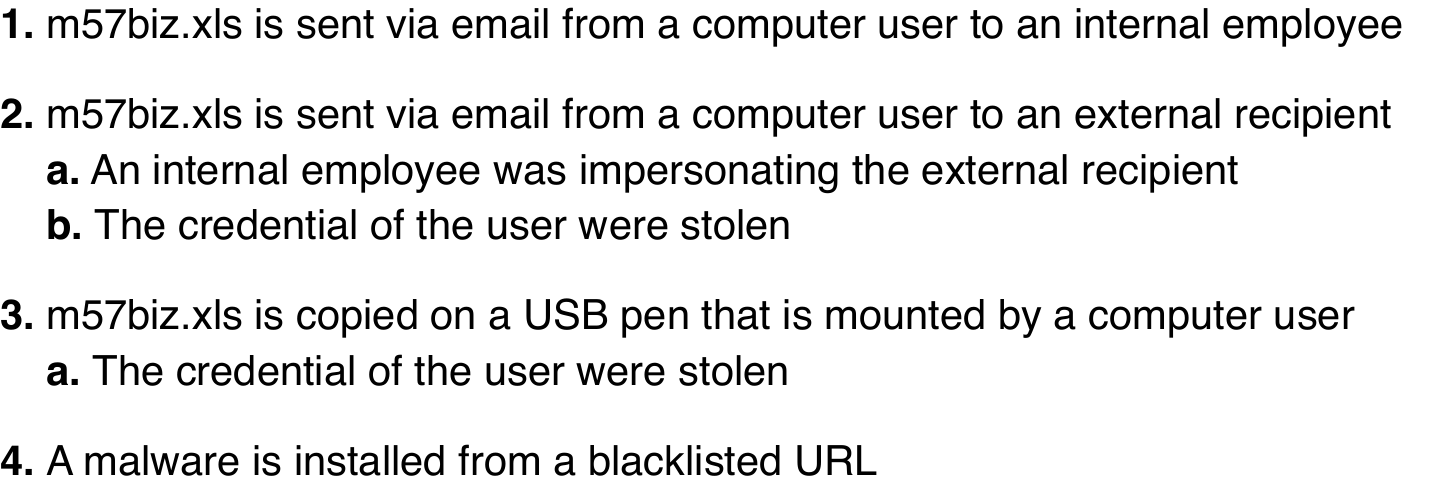}
\caption{Investigative hypotheses}\label{fig:hypotheses}
\end{figure}

\begin{figure}[htpb]
\centering
\includegraphics[width=0.98\columnwidth]{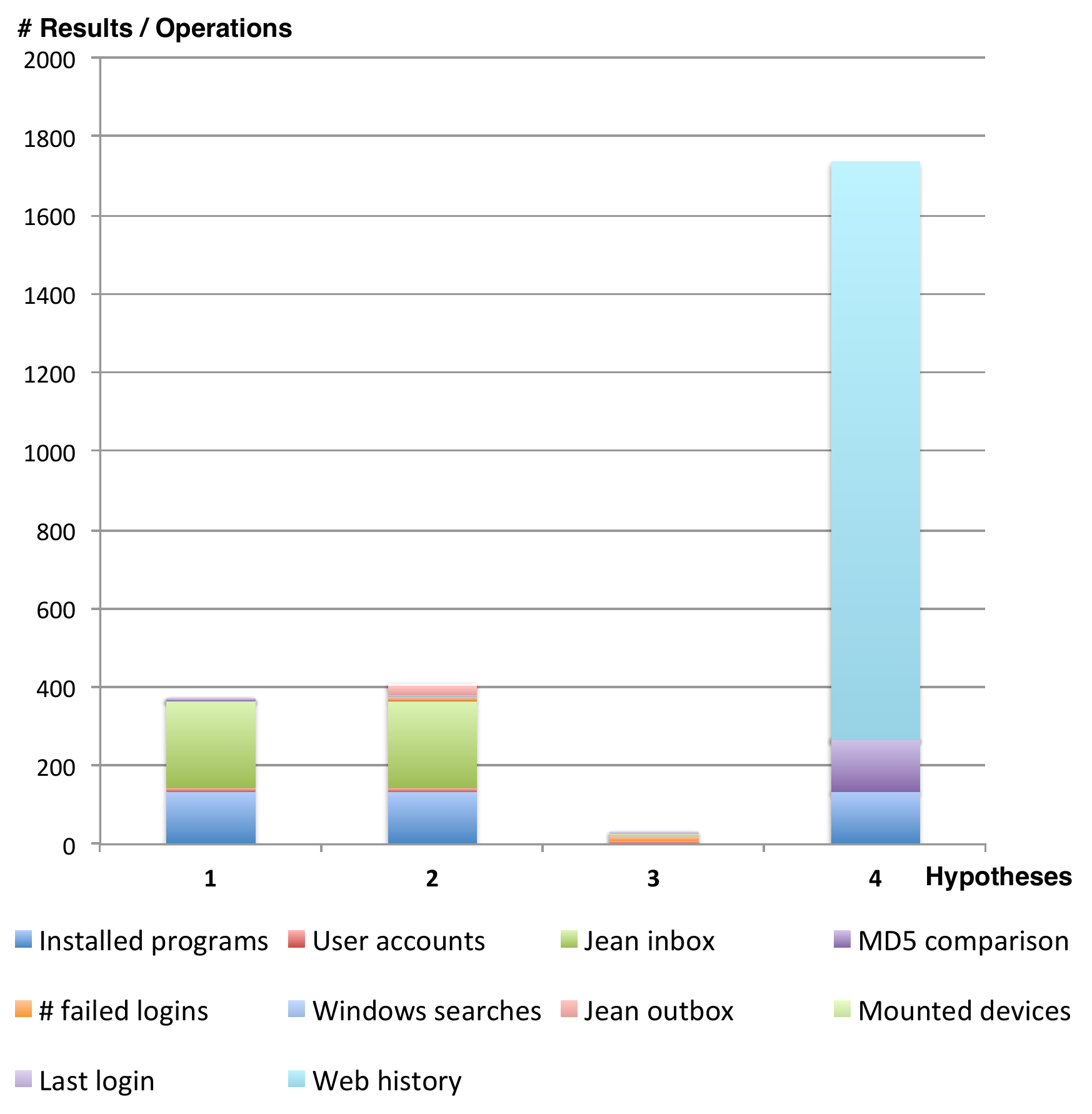}
\caption{\# Results/Operations in a Traditional Investigation.}\label{fig:traditional}
\end{figure}

We acquired the image of the computer hard drive by using the
Sleuthkit \& Autopsy tool~\cite{Sleuthkit} and we discovered that the
installed operating system is Windows XP Service Pack 3. We also used
PSTViewer~\cite{PSTViewer} to analyse sent/received emails and
RegRipper~\cite{RegRipper} to analyse Windows Registry hives and
WinMD5~\cite{WinMD5} to compute file hashes. By searching for one of
the strings contained in the confidential document ({\tt
  \$1,009,000}), the keyword search functionality of Autopsy only gave
one result ({\tt C:/.../m57biz.xls}). This document was created on
{\tt 2008-06-16T16:13:51} and was last accessed on {\tt
  2008-07-20T01:28:03} by Jean. All the hypotheses we formulated for
this case are shown in Figure~\ref{fig:hypotheses} and the amount of
evidence inspected (coming from search results) and operations
performed for each hypothesis is shown in
Figure~\ref{fig:traditional}.

In the first hypothesis we speculate that the
document is sent as an email attachment to an internal employee. Since
Outlook is the only email client in the list of the installed programs (133), we analysed available Outlook data files ({\tt
  admin\-istra\-tor.pst} and {\tt out\-look.pst})  containing  Jean inbox
(222 received emails) and outbox (23 sent emails), and the
administrator inbox (1 received emails) and outbox (0 sent emails). We
noticed that the confidential document was sent as an email attachment
by Jean as a response to another email that she received from an
external address ({\tt tuckgor\-ge@gmail.com}). From this evidence we
speculate that Jean was the victim of a phishing attack and hypothesis
1 is not viable.

Hence, the second hypothesis we formulated is that the document is sent by Jean as an email attachment to an
external email address. We also formulated additional explanations -
sub-hypothe\-ses. We thought that Jean simulated to be the victim of a phishing
attack, i.e.  phishing emails were coming from her laptop (2.a).
 However, from the evidence available it was
not possible to understand if Jean was impersonating the sender of the
phishing email. We also formulated an additional explanation 
that Jean's credentials were stolen (2.b). To demonstrate this sub-hypothesis we analysed the
sam Windows registry hive to identify created users (10 users), their
last login time and the number of failed logins. Among the users, only
the Administrator (last login on {\tt 2008-07-21T01\-:22:18} 21 July 2008 01:22:18), Jean (last login
on {\tt 2008-07-2\-0T00:00\-:41}) and Devon (last login on {\tt 2008-07-12T03:02:\-47})
 logged onto the laptop at least once and no failed login attempt was made. 
Given the identified evidence, we do not have enough elements to
demonstrate sub-hypothesis 2.b.

The third hypothesis we formulated is that the confidential document
was copied by a user on a storage device. To demonstrate this hypothesis we
inspected the system Windows registry hive and we noticed that among
the removable media (3 results) a USB pen ({\tt S/N: 7\&162a4319\&0})
was mounted on {\tt E:} on {\tt 2008-07-20\-T01:26\-:18}. Although the
confidential document was last accessed near after  this USB storage was mounted, we cannot prove that the
confidential document was copied onto a USB storage.

The fourth hypothesis we formulated is that a malware was downloaded form a blacklisted URL. 
However, only one of the installed programs ({\tt QQBubble\-Arena})
can be malicious, since its signature does not belong to the NIST National Software
Reference Library (NSRL)~\cite{NIST}, which contains the files that are known to
be good. From the web history ($\sim$1470 results) we did not
identify any blacklisted URLs. Indeed,  we can only conclude that
hypotheses 2 and 3 are likely. 


\subsection{Reactive Digital Investigation}
\label{sec:react-invest}

As a
first step we modeled the crime scene associated with the digital forensic
scenario. This is very similar to
the one proposed for our previous case study. The general model 
also includes additional entities to represent \texttt{url}s,
\texttt{program}s, more specific programs such as email clients
(\texttt{eClient}) and \texttt{browser}s, \texttt{email} accounts and
files \texttt{hash}es. 
Programs can also have a state as
they can be \texttt{Installed}. We also included in the
general model a function that identifies the signature of a
file (\texttt{MD5}) and two untimed predicates that indicate if a program is not a
malware (\texttt{InNSRL}) and if a url is blacklisted (\texttt{IsBlack}). The
concrete model includes additional monitored events to represent
sent emails (\texttt{Sent\_Email}), sent emails with attachments (\texttt{Sent\_Attach}), file
searches (\texttt{Fi\-le\_Search}) and web requests
(\texttt{Web\_Request}) for urls.
As an investigation  starting point we considered the time instant in which the
confidential file was created. All programs that were installed before
that time were initially considered as installed. At that point
in time all the users were
considered not logged and all devices not mounted.

\begin{figure}[htpb]
\centering
\includegraphics[width=0.98\columnwidth]{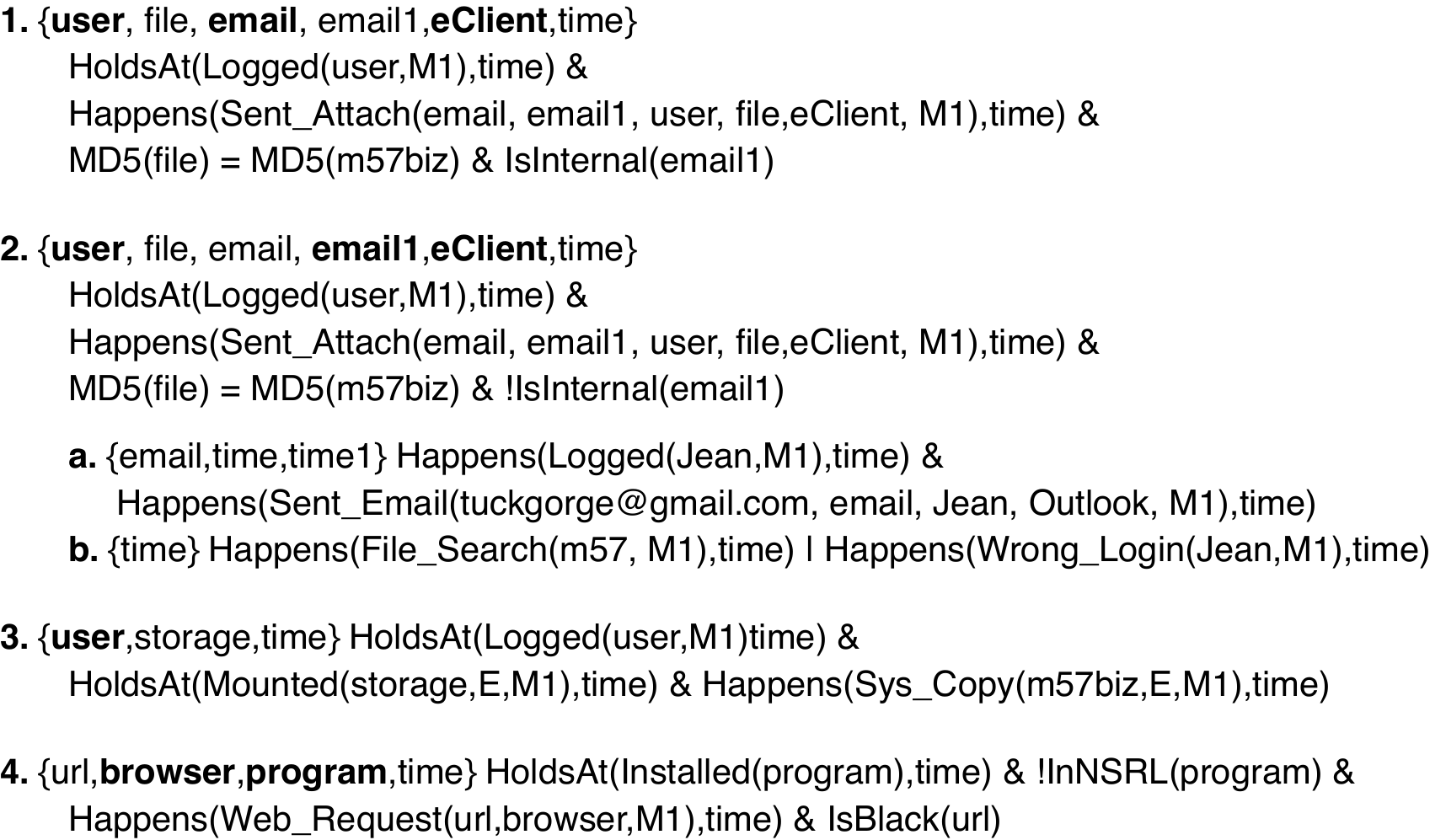}
\caption{Generic hypotheses of the evaluation scenario.}\label{fig:hExperimental}
\end{figure}

The hypotheses of a crime were modelled as shown in
Figure~\ref{fig:hExperimental}. Each hypothesis was customised
depending on the elements represented in bold. For example, hypotheses
1 and 2 are customised depending on the users (\texttt{Jean},
\texttt{Devon}, and \texttt{Administra\-tor}), on the email accounts
found in the disk image (e.g., \texttt{jean@\-m57.\-biz},
\texttt{alison@m57.biz}) (6 accounts) and on the
email clients (\texttt{Outlook}), for a total of 18 options.
Hypothesis 3 is customised depending on the users (3 options), while
hypothesis 4 is customised depending on the installed programs (133)
and the browsers (\texttt{Mozilla} and \texttt{IE}), for a total of 266 options.

All the evidence that is possible to gather from the disk image was
converted into an ordered sequence of events.  Sub-hypotheses were
only evaluated for the concrete values on which their parent hypotheses are satisfied. For example, since hypothesis 2 is completely
satisfied for user \texttt{Jane}, email
\texttt{jean@m57biz.com}, and email1
\texttt{tuck\-gorge$@$gmail.com}, hypotheses 2.a and 2.b were
only evaluated for these values.  However, hypotheses 2.a and 2.b are
not satisfied.
Hypothesis 3 is 75\% likely since only one of the monitored events is
abduced (event \texttt{Sys\_Copy}). Indeed the results we obtained are consistent with those
obtained in a traditional digital investigation and are slightly more
accurate.

\begin{figure}[htpb]
\centering
\includegraphics[width=0.9\columnwidth]{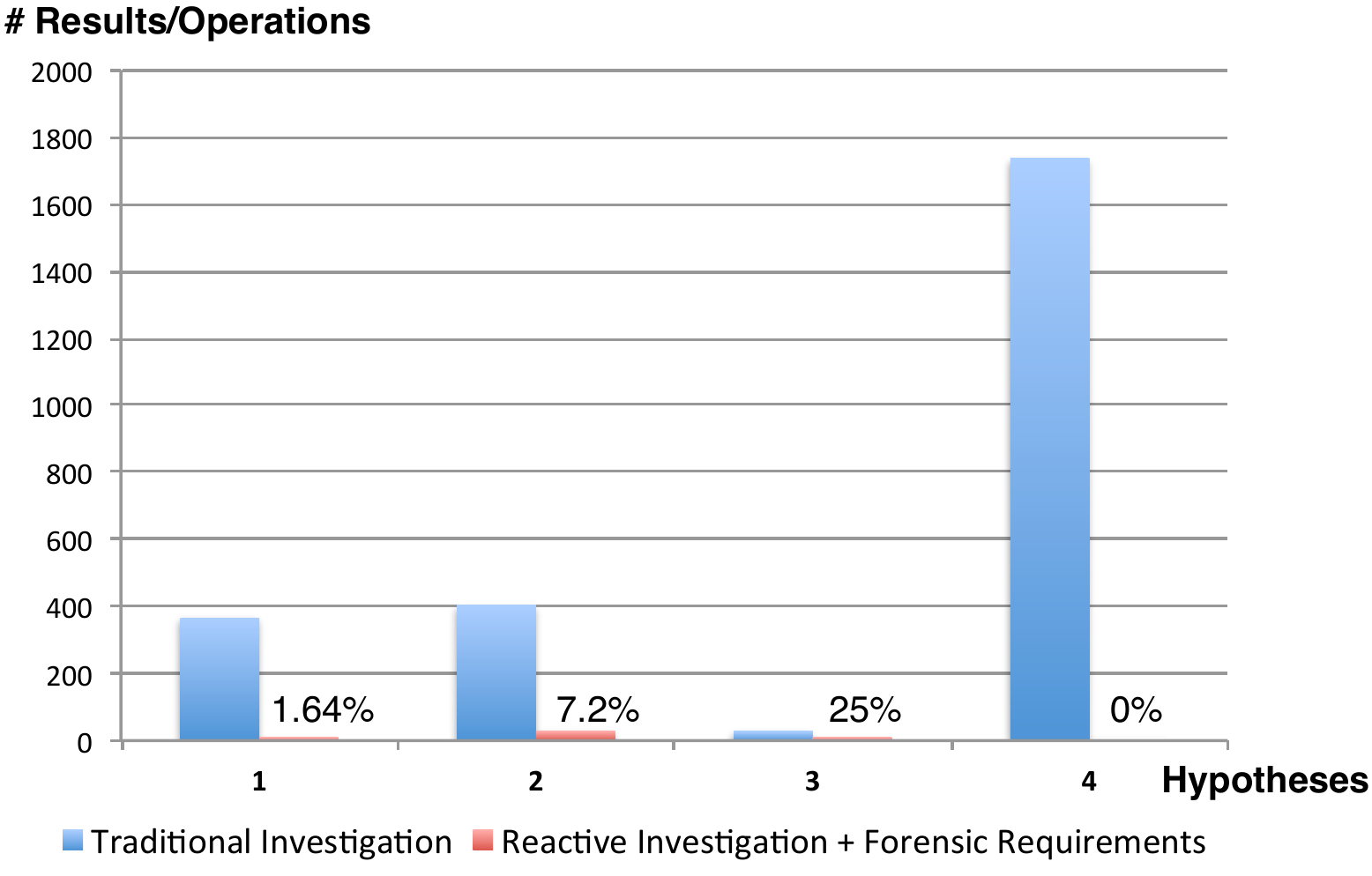}
\caption{Effectiveness of reactive activities.}\label{fig:reactive}
\end{figure}

Forensic requirements make it possible to reduce the amount of evidence
used to evaluate each hypothesis, since only the events necessary to
evaluate the predicates in the hypothesis definition are
considered. Figure~\ref{fig:reactive} shows that  the amount of
evidence considered to evaluate each hypothesis in the reactive case
was much smaller than the evidence collected in a traditional digital investigation. The percentage values
in the figure indicate the relative proportion of evidence collected
in the reactive case compared  to the traditional one. 

\subsection{Proactive \& Reactive Digital Investigation}
\label{sec:case-3}

To evaluate proactive activities we
simulated a sequence of possible monitored events that could have been collected
proactively. We envisioned two possible
situations. In the first case, we crafted an event snapshot that
complies with the evidence collected from the disk image and shows that Jean was pretending to be the victim of a phishing
attack (i.e. she was sending emails from her laptop as
\texttt{tuckgorge@gmail.com}). 
We assumed that this evidence was not
available in the disk image because it was concealed. The evaluation
of the crafted event snapshot shows that hypothesis 2.a is completely satisfied. 
In the second case, we crafted an event snapshot demonstrating
that Jean copied the confidential document onto a USB pen. In this
case, hypothesis 3 is completely satisfied, since event
\texttt{Sys\_Copy} is considered in the evaluation of the hypothesis
and does not
have to be abducted.
The aforementioned cases demonstrate the usefulness of
proactive investigations when some evidence is concealed by an
offender (case 1) or is ephemeral and cannot be retrieved from a disk image
(case 2).

\begin{figure}[htpb]
\centering
\includegraphics[width=0.8\columnwidth]{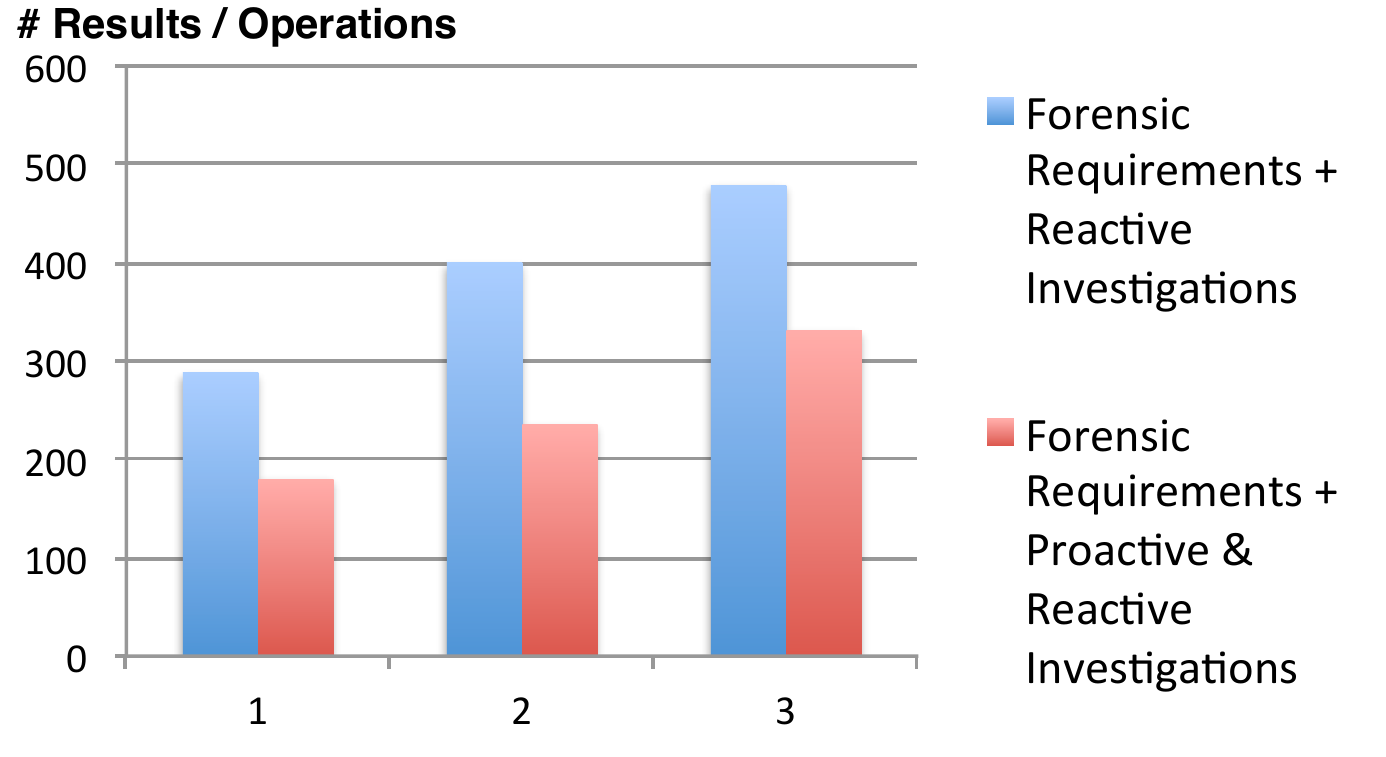}
\caption{Effectiveness of proactive activities.}\label{fig:proactive}
\end{figure}

We estimated the effectiveness of proactive activities by focusing on case 1. Together with the crafted event snapshot for hypothesis 2.a, we also
included 3 event
snapshots that did not cause the satisfaction of any hypothesis. In
addition, we also assumed a new monitored event is triggered hourly, with a uniform distribution. Note that
we considered that  hypothesis 2.a is associated with a suspicious event condition
that a user is logged on the computer. Indeed events \texttt{Sys\_Login} and
\texttt{Sys\_Logout} will be collected to check the start and stop
condition and, additionally, events \texttt{Sent\_Email} and
\texttt{Sent\_Attach} will be gathered during the full evidence collection.
For this set of monitored
events  the Proactive Collection only
preserves the snapshots related to the satisfaction of a suspicious
events and avoids gathering all the events necessary to
satisfy a hypothesis. The overhead of the Proactive Collection is only given by the 
the events necessary to evaluate the start and stop
conditions. Proactive activities provide non negligible savings in the
amount of evidence analysed, as shown in Figure~\ref{fig:proactive}.

The storage space required to preserve evidence collected proactively
for this case can be inferred by examining the workloads of local systems within an organization. We expect that $\sim$10 GB per desktop system per year would be a reasonable internal size~\cite{Shields.DI.2011}.



\subsection{Discussion}
This section discusses the limitations of our approach and proposes
possible mitigation actions. 
Our approach makes the assumption that the possible crimes that can be
committed can be defined in advance. 
For this reason, it is not viable
to investigate those crimes that exploit undiscovered vulnerabilities,
or that can be performed by unforeseen offenders, or that simply do
not follow a predefined scheme (e.g., intrusions, DDoS). However, in
the future we are planning to investigate how learning techniques can
be employed to integrate the results of completed investigations or
new discovered vulnerabilities and attacks in the model of forensic requirements.
Furthermore, our approach does not aim to
completely replace traditional digital investigations, since we
recognise that in same cases the random guess by an experienced investigator
can be very close to the reality. Instead, we
aim to speedup repetitive analysis and collection activities that are
more suitable to 
be automated.

Modelling the hypotheses can be cumbersome and
error-prone and requires some automation. Our work is amenable to
leveraging existing  model-based diagnosis approaches (e.g.,
\cite{Elsaesser.2001,Keppens.AIL.2003,Keppens.CBM.2011}) that generate speculative hypotheses of a crime
by using, for example, planning techniques. In particular, in the
future we might encode the hypotheses generation as a planning problem that
starts from the initial state of the crime scene and terminates when
the crime condition is achieved.

The Event Calculus-based analysis has an exponential complexity. 
For this reason, the Event Calculus representing the potential
hypotheses must be built in a way that the problem dimension does not exceed
$\sim$10000 variables~\cite{Mueller.2006}, which is the saturation
point of SAT based problems. To achieve this objective  we
employed some heuristics. A heuristic tries to decompose a hypothesis into
sub-hypotheses whose predicates  to be evaluated are disjunct and temporally independent. A similar
strategy was employed for 
hypotheses 2 and 2.a.  Another heuristic reduces the length of
the temporal window by splitting a hypothesis into 2 or more
hypotheses evaluated over shorter and adjacent temporal windows. This strategy can
be suitable for situations in which a
sequence of monitored events leads the crime scene again to the
condition in which a hypothesis was initially evaluated.

\section{Related Work}
\label{sec:relWork}

We now review related work on automating evidence collection, analysis, and presentation.
\subsection{Evidence Collection}
Existing tools that perform evidence collection are aimed to support computer forensics~\cite{GuidanceSoftware,ArxSys,SpectorSoft,Sleuthkit}, data recovery~\cite{FTK} and live data collection~\cite{IEF,Helix}. 
However, these approaches do not apply proactive evidence collection and analysis during the normal functioning of a system. Live acquisition of digital evidence is only performed after an incident happened and, therefore, some evidence or the traces concealed by an offender can be missed. Shield et al.~\cite{Shields.DI.2011} have recently proposed the idea to perform continuous proactive evidence collection. However, they do not suggest any possible way to select the evidence to be collected depending on the crimes that can take place in a certain environment. 
Compared to previous work, our approach does not only preserve important evidence that could be lost, but it also allows using collected evidence to trigger and guide an investigation, by indicating hypotheses to examine and additional evidence to collect.

\subsection{Evidence Analysis}
Formal techniques have been mainly used to analyse digital
evidence. Petri Nets were used to model occurred events and identify
the root causes that allowed an incident to
occur~\cite{Stephenson.DI.2003}. Gladyshev and
Patel~\cite{Gladyshev.DI.2004} formalise evidence as a series of
witness stories that restrict the possible computation of a finite
state machine that describes the behaviour of the system. All possible
scenarios of incident are obtained by backtracing transitions from the
initial state of the system.  Other work is specialised on identifying
attackers' traces (e.g., evidence and timestamps improperly
manipulated by an attacker), from violations of invariant
relationships between digital objects~\cite{Stallard.ACSAC.2003} or by
applying model checking techniques on a set of events expressed in a
multi-sorted algebra~\cite{Arasteh.DI.2007}.

Model-based reasoning
techniques~\cite{Elsaesser.2001,Keppens.AIL.2003,Keppens.CBM.2011,rekhis.CS.2005} have
been used extensively to deduce investigative hypotheses. Plan recognition techniques have been applied to
generate possible attacks sequences, simulate them on the victim model, and
perform pattern matching recognition between their side-effects and
log files entries. Abductive reasoning~\cite{Keppens.AIL.2003} was
used to explain a crime scenario from
a set of events and domain assumptions and to suggest additional
evidence to be collected. Experts' assessments of the probability of
the hypotheses  was also used to rank the evidence collection
strategies to  be performed during an investigation~\cite{Keppens.CBM.2011}.
However, the aforementioned approaches are reactive and available
evidence might not always be enough to reconstruct a crime. For this
reason, Rekhis et al.~\cite{rekhis.CS.2005} propose a language to create
hypothetical attacks scenarios in case no attack can be
explained. Similarly to model-based approaches, our work allows reasoning on
speculative hypotheses that are implicitly inferred from collected evidence~\cite{Elsaesser.2001,Keppens.AIL.2003,Keppens.CBM.2011} or explicitly
modeled~\cite{rekhis.CS.2005}. However, our approach is not focused on proposing a new
technique for evidence analysis. Our objective, instead is to use
forensic requirements to engineer systems able to preserve important evidence that can be
fundamental to solve a case. When an investigation starts the amount
of evidence analysed is highly reduced compared to the case when all
the evidence is collected. Despite the idea of integrating proactive
collection and analysis within a digital investigation has been
already proposed, existing work~\cite{Grobler.ARES.2010,Alharbi.ICISA.2011} do not offer a pathway to
implementation and do not specify how proactive and reactive
activities can be
coordinated.

\subsection{Presentation}
Our work uses structured arguments to demonstrate how collected evidence can demonstrate a set of hypotheses to be defended in court. Another work~\cite{Ernes} proposes to achieve the same objective by testing and replaying an attack in an isolated environment that is similar to the real one. When the test is finished, the analyst can relate the effects of the attack in the virtual environment to the digital evidence in the digital crime scene. If the identified effects do not support the hypotheses, the hypotheses should be reformulated, and the necessary test events should be replayed. The main drawback of this approach is that it requires reproducing an attack in a realistic environment to understand how it took place. Instead, we explicitly model the crime scene and the hypotheses of a crime to build structured arguments that formally demonstrate a hypothesis and explain how a crime took place.

\section{Conclusions and Future Work}
\label{sec:conclusions}

This
paper has proposed an approach to engineer forensic-ready sofware
systems. We proposed an adaptive  process to systematically perform the activities
to be conducted before and during a
digital forensic investigation. The process
performs proactive activities to preserve important evidence and
suggests immediate investigative directions. Second, we introduced the notion of forensics requirements to systematically
configure the activities of the proposed digital forensics process depending on a
specific crime scene and on the speculative hypotheses of a crime.
Finally, we explained how structured arguments can be used to present
the findings of an investigation. Our results suggest that our approach reduces significantly the evidence that needs to be collected and  the hypotheses that need to be analysed during an investigation. 
In the future we are planning to apply our approach in pervasive and distributed environments, such as cloud platforms, where a large amount of evidence is ephemeral.  We are also investigating new techniques to elicit forensics requirements from existing regulations.


\bibliographystyle{abbrv}
\bibliography{sig-alternate}  

\begin{thebibliography}{10}

\bibitem{FTK}
{AccessData}.
\newblock {FTK -- AccessData Digital Forensics Software}.
\newblock \url{http://www.accessdata.com/products/digital-forensics/ftk}.

\bibitem{Alharbi.ICISA.2011}
S.~Alharbi, J.~H. Weber-Jahnke, and I.~Traor\'e.
\newblock {The Proactive and Reactive Digital Forensics Investigation Process:
  A Systematic Literature Review}.
\newblock In {\em Proc. of the 5th Int. Conf. on Information Security and
  Assurance}, pages 87--100, 2011.

\bibitem{Arasteh.DI.2007}
A.~R. Arasteh, M.~Debbabi, A.~Sakha, and M.~Saleh.
\newblock {Analyzing Multiple Logs for Forensics Evidence}.
\newblock In {\em Digital Investigations}, volume~4, pages 82--91, 2007.

\bibitem{ArxSys}
{ArxSys}.
\newblock {Digital Forensics Framework}.
\newblock \url{http://www.digital-forensic.org}.

\bibitem{axelsson2000intrusion}
S.~Axelsson.
\newblock Intrusion detection systems: A survey and taxonomy.
\newblock Technical report, Chalmers University of Technology, 2000.

\bibitem{Sleuthkit}
B.~Carrier.
\newblock {The Sleuth Kit (TSK) \& Autopsy: Open Source Digital Forensic
  Tools}.
\newblock \url{http://www.sleuthkit.org}.

\bibitem{Carrier.DFRW.2004}
B.~Carrier and E.~H. Spafford.
\newblock {An Event-Based Digital Forensic Investigation Framework}.
\newblock In {\em Proceedings of the 4th Annual Digital Forensic Research
  Workshop}, 2004.

\bibitem{RegRipper}
H.~Carvey, C.~Harrell, and B.~Shavers.
\newblock {RegRipper}.
\newblock \url{http://regripper.wordpress.com}.

\bibitem{PSTViewer}
{CNET}.
\newblock {PST Viewer}.
\newblock \url{http://download.cnet.com/PST-Viewer/3000-2369_4-75289424.html}.

\bibitem{Helix}
{e-fense}.
\newblock {Helix}.
\newblock \url{http://www.e-fense.com/products.php}.

\bibitem{Elsaesser.2001}
C.~Elsaesser and M.~C. Tanner.
\newblock Automated diagnosis for computer forensics.
\newblock In {\em The Mitre Corporation}, 2001.

\bibitem{Ernes}
A.~Ernes, P.Haas, G.~Vigna, and R.~A. Kemmerer.
\newblock {Digital Forensic Reconstruction and the Virtual Security Testbed
  {ViSe}}.
\newblock In {\em Proceedings of the 3rd International Conference on Detection
  of Intrusions and Malware \& Vulnerability Assessment}, pages 144--163.
  Springer, 2006.

\bibitem{Franqueira.RE.2011}
V.~N.~L. Franqueira, T.~T. Tun, Y.~Yu, R.~Wieringa, and B.~Nuseibeh.
\newblock {Risk and Argument: A Risk-Based Argumentation Method for Practical
  Security}.
\newblock In {\em Proceedings of the 19th International Requirements
  Engineering Conference}, pages 239--248, 2011.

\bibitem{Freeman.IL.2008}
J.~B. Freeman.
\newblock {Argument Strength, the Toulmin Model, and Ampliative Probability}.
\newblock {\em Informal Logic}, 25(40), 2008.

\bibitem{DigitalCorpora}
S.~Garfinkel.
\newblock Digital corpora, m57-jean scenario.
\newblock \url{http://digitalcorpora.org/corpora/scenarios/m57-jean}, 2013.

\bibitem{Garfinkel.DFRWS.2009}
S.~Garfinkel, P.~Farrell, V.~Roussev, and G.~Dinolt.
\newblock {Bringing Science to Digital Forensics with Standardized Forensic
  Corpora}.
\newblock In {\em Proceedings of the 9th Annual Digital Forensic Research
  Workshop}, pages 2--11, 2009.

\bibitem{Giannakopoulou.2003}
D.~Giannakopoulou and J.~Magee.
\newblock {Fluent Model Checking for Event-Based Systems}.
\newblock In {\em Proceedings of the 9th European Software Engineering
  Conference}, pages 257--266. ACM, 2003.

\bibitem{Gladyshev.DI.2004}
P.~Gladyshev and A.~Patel.
\newblock {Finite State Machine Approach to Digital Event Reconstruction}.
\newblock {\em Digital Investigations}, 1:130--149, 2004.

\bibitem{Grobler.ARES.2010}
T.~Grobler, C.~P. Louwrens, and S.~H. von Solms.
\newblock {A Framework to Guide the Implementation of Proactive Digital
  Forensics in Organisations}.
\newblock In {\em Proceedings of the 5th International Conference on
  Availability, Reliability and Security}, pages 677--682, 2010.

\bibitem{GuidanceSoftware}
{Guidance Software}.
\newblock {EnCase Forensics -- Computer Forensics Data Collection for Digital
  Evidence Examiners}.
\newblock \url{http://www.guidancesoftare.com/encase-forensics.htm}.

\bibitem{Haley.TSE.2008}
C.~Haley, R.~Laney, J.~Moffett, and B.~Nuseibeh.
\newblock {Security Requirements Engineering: A Framework for Representation
  and Analysis}.
\newblock {\em IEEE Transactions on Software Engineering}, 34(1):133--153,
  2008.

\bibitem{Ingolfo.REFSQ.2013}
S.~Ingolfo, A.~Siena, I.~Jureta, A.~Susi, A.~Perini, and J.~Mylopoulos.
\newblock {Choosing Compliance Solutions through Stakeholder Preferences}.
\newblock In {\em Proc. of the 19th Working Conference on Requirements
  Engineering: Foundation for Software Quality}, pages 206--220. Springer,
  2013.

\bibitem{Keppens.CBM.2011}
J.~Keppens, Q.~Shen, and C.~Price.
\newblock {Compositional Bayesian modelling for computation of evidence
  collection strategies}.
\newblock {\em {Applied Intelligence}}, 35(1):134--161, 2011.

\bibitem{Keppens.AIL.2003}
J.~Keppens and J.~Zeleznikow.
\newblock {A Model Based Reasoning Approach for Generating Plausible Crime
  Scenarios from Evidence}.
\newblock In {\em Proceedings of the 9th International Conference on Artificial
  Intelligence and Law}, pages 51--59. ACM, 2003.

\bibitem{IEF}
{MAGNET FORENSICS}.
\newblock {Internet Evidence Finder}.

\bibitem{Decreasoner}
E.~T. Mueller.
\newblock {Discrete Event Calculus Reasoner Documentation}.
\newblock \url{http://decreasoner.sourceforge.net/csr/decreasoner.pdf}.

\bibitem{Mueller.2006}
E.~T. Mueller.
\newblock {\em {Commonsense Reasoning}}.
\newblock Morgan Kaufmann, 2006.

\bibitem{Beebe.DI.2005}
J.~G.~C. N.~Beebe.
\newblock {A Hierarchical, Objectives-Based Framework for the Digital
  Investigations Proces}.
\newblock {\em Digital Investigations}, 2:147--167, 2005.

\bibitem{NIST}
{National Institute of Standards and Technology (NIST)}.
\newblock {National Software Reference Library}.
\newblock \url{http://www.nsrl.nist.gov}.

\bibitem{NewYorkTimes}
{New York Times}.
\newblock {Raj Rajaratnam's Galleon Group Founder Convicted in Insider Trading
  Case}.
\newblock \url{topics.nytimes.com/top/reference/timestopics/people/r/
  raj\_rajaratnam/index.html}.

\bibitem{Palmer.DFRWS.2001}
G.~Palmer.
\newblock {A Road Map for Digital Forensics Research}.
\newblock Technical report, Air Force Research Lab, Rome, 2001.

\bibitem{Evaluation}
L.~Pasquale.
\newblock {Engineering Adaptive Digital Investigations Using Forensic
  Requirements: Experimental Results}.
\newblock \url{http://staff.lero.ie/lpasqua/files/2013/09/Evaluation.pdf},
  2013.

\bibitem{Pasquale.RE.2013}
L.~Pasquale, Y.~Yu, M.~Salehie, L.~Cavallaro, T.~T. Tun, and B.~Nuseibeh.
\newblock {Requirements-Driven Adaptive Digital Forensics}.
\newblock In {\em Proceedings of the 21st Internatioanl Requirements
  Engineering Conference}, 2013.
\newblock {Poster Track, to appear}.

\bibitem{Pollitt.DF.2007}
M.~M. Pollitt.
\newblock {An Ad Hoc Review of Digital Forensics Models}.
\newblock In {\em Proceedings of the 2nd International Workshop on Systematic
  Approaches to Digital Forensics Engineering}, pages 43--54, 2007.

\bibitem{rekhis.CS.2005}
S.~Rekhis and N.~Boudriga.
\newblock A temporal logic-based model for forensic investigation in networked
  system security.
\newblock In {\em Computer Network Security}, pages 325--338. Springer, 2005.

\bibitem{Rowlingson.IJDE.2004}
R.~Rowlingson.
\newblock {A Ten Step Process for Forensic Readiness}.
\newblock {\em International Journal of Digital Evidence}, 2(3):1--28, 2004.

\bibitem{Shields.DI.2011}
C.~Shields, O.~Frieder, and M.~Maloof.
\newblock {A System for the Proactive, Continuous, and Efficient Collection of
  Digital Forensic Evidence}.
\newblock In {\em Digital Investigations}, volume~8, pages 3--13, 2011.

\bibitem{SpectorSoft}
{SpectorSoft}.
\newblock {Investigate Employee Computer Activity with SPECTOR CNE}.
\newblock \url{http://www.spectorcne.com/}.

\bibitem{Stallard.ACSAC.2003}
T.~Stallard and K.~N. Levitt.
\newblock {Automated Analysis for Digital Forensics Science: Semantic Integrity
  Checking}.
\newblock In {\em Proc. of the Annual Computer Security Application
  Conference}, pages 160--167, 2003.

\bibitem{Stephenson.DI.2003}
P.~Stephenson.
\newblock {Modeling of Post-Incident Root Cause Analysis}.
\newblock {\em International Journal of Digital Evidence}, 2, 2003.

\bibitem{FBICrimeReport}
{The Federal Bureau of Investigation (FBI)}.
\newblock {Financial Crimes Report 2010-2011}.
\newblock
  \url{http://www.fbi.gov/stats-services/publications/financial-crimes-report-2010-2011},
  2011.

\bibitem{ISO}
{The ISO 27000 Directory}.
\newblock {An Introduction to ISO 27001, ISO 27002...ISO 27008}.
\newblock \url{http://www.27000.org}.

\bibitem{Toulmin.2003}
S.~E. Toulmin.
\newblock {\em {The Uses of Arguments}}.
\newblock Cambridge University Press, 2003.

\bibitem{Tun.RE.2012}
T.~T. Tun, A.~K. Bandara, B.~A. Price, Y.~Yu, C.~B. Haley, I.~Omoronyia, and
  B.~Nuseibeh.
\newblock {Privacy Arguments: Analysing Selective Disclosure Requirements for
  Mobile Applications}.
\newblock In {\em Proceedings of the 20th International Requirements
  Engineering Conference}, pages 131--140, 2012.

\bibitem{Willassen.SAC.2008}
S.~Y. Willassen.
\newblock {Using Simplified Event Calculus in Digital Investigation}.
\newblock In {\em Proc. of the ACM Symposium on Applied Computing}, pages
  1438--1442, 2008.

\bibitem{WinMD5}
{WinMD5.com}.
\newblock {WinMD5Free}.
\newblock \url{http://www.winmd5.com}.

\end{thebibliography}
\end{document}